%% file: main.tex
\documentclass[sigconf, twocolumn]{acmart}
\usepackage{balance}
\usepackage{enumitem}
\AtBeginDocument{%
  \providecommand\BibTeX{{%
    \normalfont B\kern-0.5em{\scshape i\kern-0.25em b}\kern-0.8em\TeX}}}

 \acmConference[]{To Appear in ACM Transactions on Cyber-Physical Systems}{ }{ }

\acmBooktitle{To Appear at ACM Transactions of Cyber Physical Systems}

\usepackage{amsmath}
\usepackage{appendix}
\usepackage{amsfonts}
\usepackage{graphicx}
\usepackage{xcolor}
\usepackage{bbm}
\usepackage[ruled,vlined,linesnumbered]{algorithm2e}
\SetArgSty{textnormal}
\usepackage[normalem]{ulem}
\usepackage[ruled,vlined]{algorithm2e}
\usepackage{caption}
\usepackage{subcaption}
\usepackage{soul}
\usepackage{mathtools}
\usepackage{pgfplots}
\pgfplotsset{compat=1.8}

\usepgfplotslibrary{statistics}
\makeatletter
\pgfplotsset{
    boxplot/hide outliers/.code={
        \def\pgfplotsplothandlerboxplot@outlier{}%
    }
}
\makeatother

\usepackage{color, colortbl}
\usepackage{bbold}

\DeclareMathOperator*{\argmax}{arg\,max}

\setcopyright{none}
\renewcommand\footnotetextcopyrightpermission[1]{}
\begin{document}

\title[Hierarchical Planning for Dynamic Resource Allocation in Smart and Connected Communities]{Hierarchical Planning for Dynamic Resource Allocation in Smart and Connected Communities}

\author{Geoffrey Pettet}
\email{geoffrey.a.pettet@vanderbilt.edu}
\affiliation{%
  \institution{Vanderbilt University}
  \city{Nashville}
  \state{TN}
  \postcode{37212}
  \country{USA}
}

\author{Ayan Mukhopadhyay}
\authornote{The author was at Stanford University when this work started and was sponsored by the Center of Automotive Research at Stanford.}
\email{ayan.mukhopadhyay@vanderbilt.edu}
\affiliation{%
  \institution{Vanderbilt University}
  \city{Nashville}
  \state{TN}
  \postcode{37212}
  \country{USA}
}

\author{Mykel J. Kochenderfer}
\email{mykel@stanford.edu}
\affiliation{%
  \institution{Stanford University}
  \city{Stanford}
  \state{CA}
  \postcode{94305}
  \country{USA}
}
\author{Abhishek Dubey}
\email{abhishek.dubey@vanderbilt.edu}
\affiliation{%
  \institution{Vanderbilt University}
  \city{Nashville}
  \state{TN}
  \postcode{37212}
  \country{USA}
}






\thanks{This work is sponsored by The National
Science Foundation under award numbers CNS1640624 and IIS1814958 and a grant from Tennessee Department of Transportation. We also acknowledge support from Google through the Cloud Research Credits.}
\renewcommand{\shortauthors}{Pettet, et al.}


\begin{abstract}
    \input{abstract}
\end{abstract}






\maketitle
 \pagestyle{fancy}
\input{introduction}
\input{ProblemFormulation}

\input{Approach}

\input{Implementation}
\input{Experiments}

\input{related_work}
\input{Conclusion}
\balance

\bibliographystyle{ACM-Reference-Format}
\bibliography{references}

\end{document}

%% file: abstract.tex
Resource allocation under uncertainty is a classic problem in city-scale cyber-physical systems. Consider emergency response, where urban planners and first responders optimize the location of ambulances to minimize expected response times to incidents such as road accidents. Typically, such problems involve sequential decision making under uncertainty and can be modeled as Markov (or semi-Markov) decision processes. The goal of the decision maker is to learn a mapping from states to actions that can maximize expected rewards. While online, offline, and decentralized approaches have been proposed to tackle such problems, scalability remains a challenge for real world use cases. We present a general approach to hierarchical planning that leverages structure in city level CPS problems for resource allocation. We use emergency response as a case study and show how a large resource allocation problem can be split into smaller problems. We then use Monte Carlo planning for solving the smaller problems and managing the interaction between them. Finally, we use data from Nashville, Tennessee, a major metropolitan area in the United States, to validate our approach. Our experiments show that the proposed approach outperforms state-of-the-art approaches used in the field of emergency response.

%% file: introduction.tex
\section{Introduction}

Dynamic resource allocation (DRA) in anticipation of uncertain demand is a common problem in city-scale cyber-physical systems (CPS)~\cite{MukhopadhyayICCPS}. In such a scenario, the decision-maker optimizes the spatial location of resources (typically called \textit{agents}) to maximize utility over time while satisfying constraints specific to the problem domain. This optimization requires the design and development of procedures that can estimate how the system will evolve and evaluate the long-term value of actions. DRA manifests in many problems at the intersection of urban management, CPS, and multi-agent systems. These include emergency response management (ERM) for ambulances and fire trucks~\cite{mukhopadhyay2020review}, allocating helper trucks~\cite{vazirizade2021learning}, designing on-demand transit~\cite{chebbi2015modeling}, and positioning electric scooters~\cite{gossling2020integrating}. All such use cases typically deal with particular incidents of interest which correspond to specific calls for service. For example, road accidents require calls for emergency responders. In anticipation of such calls, planners proactively optimize the allocation of resources. The allocation problem can be modeled as a sequential decision-making problem under uncertainty, which can be solved to maximize a domain-specific utility function. For example, maximizing the total demand that can be serviced or minimizing the expected response time to incidents are commonly used objectives~\cite{mukhopadhyay2020review}.

While we present a general approach for dynamic resource allocation in urban areas, we focus mainly on the problem of emergency response as a case study since it is a critical problem faced by communities across the globe. With road accidents accounting for 1.25 million deaths globally and 240 million emergency medical services (EMS) calls made in the USA each year, there is a critical need for a proactive and effective response to these emergencies~\cite{mukhopadhyay2020review,911stats}. In addition to responding to frequent incidents each day, emergency response management (ERM) addresses large-scale disasters due to natural hazards (climate driven disasters caused more than \$90 billion of losses in the US in 2018~\cite{severeClimate}) and man made attacks. The lack of active and timely response to emergencies constitutes a threat to human lives and has resulted in mounting costs. Minimizing the time to respond to emergency incidents is critical in emergency response. Therefore, governments and private agencies often try to proactively optimize ambulance locations while considering constraints on the number of available ambulances and locations where they can be stationed. In addition, they address the problem of resource dispatch, in which agents\footnote{The actors of decision making are referred to as agents (e.g., ambulances)} must be selected and asked to spatially move from their location to the location of the demand to address the task at hand. For example, ambulances must move to the scene of the incidents to provide assistance to patients. Effectively, ERM pipelines are an important example of human-in-the-loop CPS (H-CPS), which introduces problem specific structure and constraints.



\begin{figure}[t]
    \centering
    \includegraphics[width=\columnwidth]{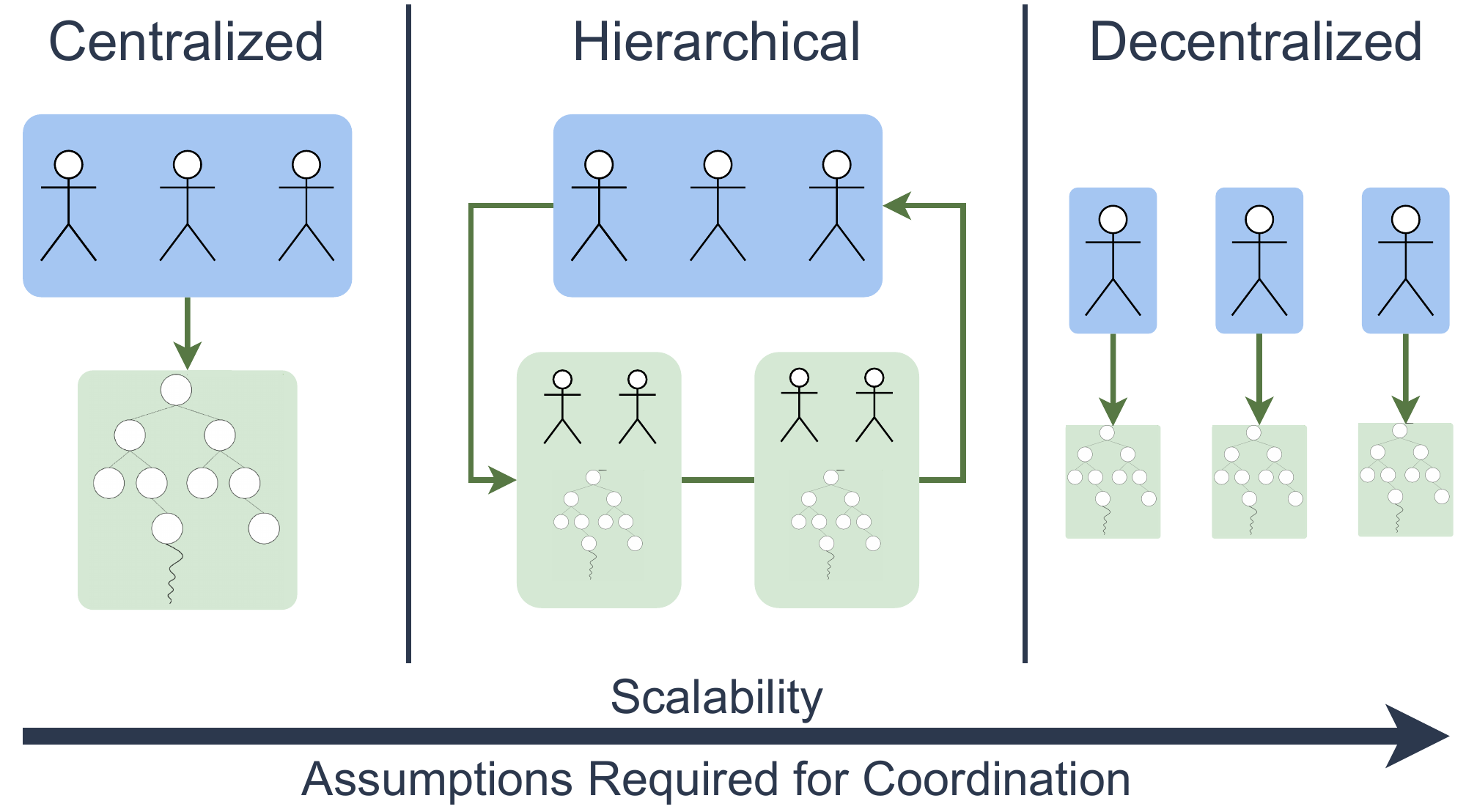}
    \caption{A spectrum of approaches for solving a dynamic resource allocation problem under uncertainty. A completely centralized approach uses a monolithic state representation. In a completely decentralized approach, each agent simulates what other agents do and performs its own action estimation. Our hierarchical approach segments the planning problem into sub-problems to improve scalability without agents estimating other agents' actions.}
    \label{fig:planning}
\end{figure}

\textbf{State of the art:} DRA and dispatch problems are typically modeled as Markov decision processes (MDP)~\cite{kochenderfer2015decision, mukhopadhyayAAMAS18, mukhopadhyay2020review}. The decision maker's goal is to find an optimal \textit{policy}, which is a mapping between system states and actions to be taken. 
There is a broad spectrum of approaches available to address resource allocation under uncertainty as shown in figure~\ref{fig:planning}. In our previous work, we applied the extremes of this continuum to ERM, with each approach having strengths and weaknesses~\cite{mukhopadhyayAAMAS18, MukhopadhyayICCPS, pettet2020algorithmic}. 


The most direct approach is to model the state of the MDP as a monolithic entity that captures the entire system in consideration, as shown on the left in figure ~\ref{fig:planning}. When real world problems are modeled as MDPs, state transitions are typically unknown. The standard method to address this issue is to use a simulator (a model of the world) to estimate an empirical distribution over the state transitions~\cite{mukhopadhyayAAMAS18}, and use the estimates to learn a policy using dynamic programming~\cite{kochenderfer2015decision}. Unfortunately, this offline approach does not scale to realistic problem settings~\cite{MukhopadhyayICCPS}. Another method is to use an online algorithm like Monte Carlo tree search (MCTS)~\cite{MukhopadhyayICCPS}. While adaptable to dynamic environments, this approach still suffers from poor scalability, taking too long to converge for realistic problem scenarios.  

On the other side of the spectrum is an entirely decentralized methodology, as shown in the extreme right in figure ~\ref{fig:planning}. In such an approach, each agent determines its course of action. As the agents cooperate to achieve a single goal, they must estimate what other agents will do in the future as they optimize their actions. For example, \citeauthor{claes2017decentralised}~\citep{claes2017decentralised} show how each agent can explore the efficacy of its actions locally by using MCTS. While such approaches are significantly more scalable than their centralized counterparts, they are sub-optimal as agents' estimates of other agents' actions can be inaccurate. Note that high fidelity models for estimating agents' actions limit scalability, and therefore decentralized approaches rely on computationally inexpensive heuristics. While decentralized approaches can be helpful in disaster scenarios where communication networks can break down, agents in urban areas (ambulances) typically have access to reliable networks, and communication is not a constraint. Therefore, approaches that ensure scalability but do not fully use available information during planning are not suitable for emergency response in urban areas, especially when fast and effective response is critical.  

We explore hierarchical planning~\cite{hauskrecht2013hierarchical}, which focuses on learning local policies, known as \textit{macros}, over subsets of the state space by leveraging the spatial structure in the problem. Our idea is motivated by the concept of \textit{jurisdictions} or \textit{action-areas} used in public policy, which create different zones to partition and better manage infrastructure. In particular, we design a principled algorithmic approach that partitions the spatial area under consideration into smaller areas. We then treat resource allocation in each resulting sub-area (called regions) as individual planning problems which are smaller than the original problem by construction. While this ensures scalability, it hurts performance because agents constrained in one region might be needed in the other region. We show how hierarchical planning can be used to facilitate the transfer of agents across regions. A top-level planner, called the inter-region planner, identifies and detects states where ``interaction'' between regions is necessary and finds actions such that the system's overall utility can be maximized. A low-level planner, called the intra-region planner, addresses allocation and dispatch within a region. A challenge with hierarchical planning is that the high-level planner must estimate rewards further along the planning pipeline to make decisions. However, such rewards can only be computed \textit{after} the low-level planner optimizes its objective function; this constraint defeats the purpose of hierarchical planning because it does not segregate the overall planning problem between two different levels. We leverage the structure of DRA and dispatch problems to address this challenge.

\input{notations}

\textbf{Contributions}: the specific contributions of this article include:

\begin{enumerate}
    \item We propose a hierarchical planning approach for resource allocation under uncertainty for smart and connected communities that scales significantly better than prior approaches. 
    
    \item We show how exogenous constraints in real world resource allocation problems can be used to naturally partition the overall decision problem into sub-problems. We create a low-level planner that focuses on finding optimal policies for the sub-problems.
    
    \item We show how a high-level planner can facilitate the exchange of resources between the sub-problems (spatial areas in our case).
    
    \item We propose two models, a data-driven surrogate model and a queue based approximation, that can estimate rewards to aid the high-level planner. In practice, responders can remain busy even after they leave the scene of the incident (e.g., ambulances transport victims of accidents to hospitals). Unlike prior work~\cite{MukhopadhyayICCPS, pettet2021hierarchical}, the data-driven approach we propose can accommodate such drop-off times.
    
    \item We use real world emergency response data from Nashville, Tennessee, a major metropolitan area in the United States, to evaluate our approach and show that it performs better than state-of-the-art approaches both in terms of efficiency, scalability, and robustness to agent failures.
    
\end{enumerate}


The paper is organized as follows. We present a general decision-theoretic formulation for spatial-temporal resource allocation under uncertainty in section~\ref{sec:problem_formulation}. Section~\ref{sec:approach} describes the overall approach, the high-level, and the low-level planner. Section \ref{sec:framework} details the implementation of our decision support system for ERM responder allocation. We present our experimental design in section~\ref{sec:exp} and the results in section~\ref{sec:exp_results}. We present related work in section~\ref{sec:related} and summarize the paper in section~\ref{sec:conclusion}. Table \ref{tab:lookup-table} summarizes the notation used throughout the paper.


%% file: notations.tex
\definecolor{Gray}{gray}{0.9}
\begin{table}[t]
\caption{Notation.}

\captionsetup{font=small}
 \resizebox{\columnwidth}{!}{%
\begin{tabular}{|ll|}
\hline
\rowcolor{Gray}
{\ul \textbf{Symbol}}     & {\ul \textbf{Definition}}                                                                       \\ \hline
$\Lambda$                 & Set of agents                                                                                   \\ \hline\rowcolor{Gray}
$D$                       & Set of depots                                                                                   \\ \hline
$\mathcal{C}(d)$          & Capacity of depot $d$                                                                           \\ \hline\rowcolor{Gray}
$G$                       & Set of cells                                                                                    \\ \hline
$R$                     & Set of regions                                                                                    \\ \hline\rowcolor{Gray}
$S$                       & State space                                                                                     \\ \hline
$A$                       & Action space                                                                                    \\ \hline\rowcolor{Gray}
$P$                       & State transition function                                                                       \\ \hline
$T$                       & Temporal transition distribution                                                                \\ \hline\rowcolor{Gray}
$\alpha$                  & Discount factor                                                                                 \\ \hline
$\rho(s, a)$                    & Reward function given action $a$ taken in state $s$                                       \\ \hline\rowcolor{Gray}
$\mathcal{A}$             & Joint agent action space                                                                        \\ \hline
$\mathcal{T}$             & Termination scheme                                                                              \\ \hline\rowcolor{Gray}
$s^t$                     & Particular state at time $t$                                                                    \\ \hline
$I^t$                     & Set of cell indices waiting to be serviced                                                     \\ \hline\rowcolor{Gray}
$\mathcal{Q}(\Lambda)$    & Set of agent state information                                                               \\ \hline
$p^t_j$                   & Position of agent $j$                                                                           \\ \hline\rowcolor{Gray}
$g^t_j$ & Destination of agent $j$                                                                                            \\ \hline
$u^t_j$ & Current status of agent $j$                                                                                      \\ \hline\rowcolor{Gray}
$s_i, s_j$                & Individual states                                                                               \\ \hline
$\sigma$                  & Action recommendation set                                                                        \\ \hline\rowcolor{Gray}
$\eta$                    & Service rate                                                                                     \\ \hline
$\gamma_g$                & Incident rate at cell $g$                                                                       \\ \hline\rowcolor{Gray}
$t_h$                     & Time since beginning of planning horizon                                                        \\ \hline
$t_r(s,a)$                   & Response time to an incident given action $a$ in state $s$                                      \\ \hline\rowcolor{Gray}
$p_j$                     & Number of agents assigned to region $r_j$                                                       \\ \hline
$\gamma_j$               & Total incident rate in region $r_j$                                                               \\ \hline\rowcolor{Gray}
$w_j(p_j, \gamma_j)$                     & Expected waiting time for incidents in region $r_j$                                \\ \hline
$D_j$               & Set of depots in region $r_j$                                                               \\ \hline\rowcolor{Gray}
$G_j$               & Set of cells in region $r_j$                                                               \\ \hline

\end{tabular}%
}
\label{tab:lookup-table}
\end{table}

%% file: ProblemFormulation.tex
\section{Problem Formulation}\label{sec:problem_formulation}


Resource allocation in anticipation of spatial-temporal requests is a common problem in urban areas. For example, ambulances respond to road accidents and emergency calls, helper trucks respond to vehicle failures, taxis provide transit service to customers, and fire trucks respond to urban fires. We refer to responders as \textit{agents} to be consistent with the terminology used in multi-agent systems~\cite{wooldridge2009introduction}. Once an incident is reported, agents are dispatched to the scene of the incident. The decision to select an agent to be dispatched can either be performed by a human expert (e.g., dispatching towing trucks), human-algorithm collaboration (e.g., dispatching ambulances), or completely by an algorithmic approach (e.g., dial-a-ride services). If no agent is available, the incident typically enters a waiting queue and is responded to when an agent becomes free. Each agent is typically housed at specific locations distributed throughout the spatial area under consideration (these could be fire stations or rented parking spots, for example). The number of such locations can vary by the type of incident and agent in consideration. For example, while taxis can wait at arbitrary locations or move around areas with high historical demand, ambulances are typically housed at rented parking lots or designated stations. We refer to such locations as \textit{depots}. Once an agent finishes servicing an incident (which might involve transporting the victim of an accident to a nearby hospital), it becomes available for service again. If there are no pending incidents, it is directed back to a depot. Therefore, there are two broad actions that the decision-maker can optimize: (1) which agent to dispatch once an incident occurs (dispatching action) and (2) which depots to send the agents to in anticipation of future demand (allocation action). Next, we introduce our assumptions and the notation we use for problem formulation. While we present a formulation that is broadly applicable to resource allocation problems under uncertainty in urban areas, we use emergency response as a case study to provide concrete examples and use-cases.

We begin with several assumptions on the problem structure. First, we assume that we are given a spatial map broken up into a finite collection of equally sized cells $G$, a set of agents $\Lambda$ that need to be allocated across these cells and dispatched to demand points, and a travel model that describes how the agents move throughout $G$. We also assume that we have access to a spatial-temporal model of demand over $G$, which is homogenous within each spatial cell. Our third assumption is that agent allocation is restricted to \textit{depots} $D$, which are located in a fixed subset of cells. Each depot $d \in D$ has a fixed capacity $\mathcal{C}(d)$, which is the number of agents it can accommodate. While the state space in this resource allocation problem evolves in continuous-time, it is convenient to view the dynamics as a set of finite decision-making states that evolve in discrete time. For example, an agent moving through an area continuously changes the state of the \textit{world} but presents no scope for decision-making unless an event occurs that needs a response or the planner redistributes the agents. As a result, the decision-maker only needs to find optimal actions for a subset of the state space that provides the opportunity to take actions. 

A key component of response problems is that agents physically move to the site of the request, which makes temporal transitions between states non-memoryless. This component causes the underlying stochastic process governing the system's evolution to be semi-Markovian. The dynamics of a set of agents working to achieve a common goal can be modeled as a Multi-Agent Semi-Markov Decision Process (MSMDP)~\cite{rohanimanesh2003learning}, which can be represented as the tuple $(S,\Lambda,\mathcal{A},P,T,\rho(s,a),\alpha,\mathcal{T})$, where $S$ is a finite state space, $\rho(s, a)$ represents the instantaneous reward for taking action $a$ in state $s$, $P$ is a state transition function, $T$ is the temporal distribution over transitions between states, $\alpha$ is a discount factor, and $\Lambda$ is a finite collection of agents where $\lambda_j \in \Lambda$ denotes the $j$th agent. The action space of the $j$th agent is represented by $A_j$, and $\mathcal{A} = \prod_{i=1}^{m} A_j$ represents the joint action space of all agents. We assume that the agents are cooperative and work to maximize the system's overall utility. $\mathcal{T}$ represents a termination scheme. Since agents take different actions with different times to completion, they may not all terminate at the same time~\cite{rohanimanesh2003learning}. We focus on asynchronous termination, where actions for a particular agent are chosen as and when the agent completes its last assigned action.\footnote{Different termination schemes are discussed by ~\citeauthor{rohanimanesh2003learning}~\cite{rohanimanesh2003learning}.}

\textbf{States:} A state at time $t$ is represented by $s^t$ and consists of a tuple $(I^t, \mathcal{Q}(s^t))$, where $I^t$ is a collection of cell indices that are waiting to be serviced, ordered according to the relative times of incident occurrence. $\mathcal{Q}(s^t)$ corresponds to information about the set of agents at time $t$ with $|\mathcal{Q}(s^t)| = |\Lambda_r|$. Each entry $q^t_j \in \mathcal{Q}(s^t)$ is a set $\{p^t_j,g^t_j,u^t_j, r_j^t, d_j^t\}$, where $p^t_j$ is the position of agent $\lambda_j$, $g^t_j$ is the destination cell that it is traveling to (which can be its current position), $u^t_j$ is used to encode its current status (busy or available), $r_j^t$ is the agent's assigned region, and $d_j^t$ is its assigned depot, all observed at time $t$. A diagram of the state is shown in figure \ref{fig:state} and discussed in detail in section \ref{sec:framework}. We assume that no two events occur simultaneously in our system model. In such a case, since the system model evolves in continuous time, we can add an arbitrarily small time interval to create two separate states. We overload the notation for states for convenience; an arbitrary state is denoted by $s$ when the time at which the state occurs is not required for discussion.

\textbf{Actions:}
Actions correspond to directing agents to valid cells to respond to incidents (demand) or wait at a depot. For a specific agent $\lambda_i \in \Lambda_r$, valid actions for a specific state $s$ are denoted by $A^i(s)$ (some actions are naturally invalid, for example, if an agent is at cell $k$ in the current state, any action not originating from cell $k$ is unavailable to the agent). Actions can be broadly divided into two categories: \textit{dispatching} actions which direct agents to service an active demand point, and \textit{allocation} actions which assign agents to wait in particular depots in anticipation of future demand. 

The manner in which agents are dispatched to the scene of the incidents varies with the type of incident. For example, an essential aspect of emergency response is that if any free agents are available when an incident is reported, then the nearest one must be greedily dispatched to attend to the incident. This constraint is a direct consequence of the bounds within which emergency responders operate and the critical nature of such incidents~\cite{mukhopadhyay2020review, mukhopadhyay2020designing, pettet2020algorithmic}. On the other hand, taxis can optimize dispatch based on long-term rewards. We focus on emergency response in our experiments and validate the approach using data collected from ambulances. As a result, the problem we consider focuses on proactively redistributing agents across a spatial area under future demand uncertainty. Nonetheless, dispatch actions are still necessary to model since they are the foundation of our reward function.


\textbf{Transitions:}
The resource allocation system model evolves through several stochastic processes. First, incidents occur at different points in time and space governed by some arrival distribution. We assume that the number of incidents in a cell $r_j \in R$ per unit time can be approximated by a Poisson distribution with mean rate $\gamma_j$ (per unit time), a commonly used model for spatial-temporal incident occurrence~\cite{mukhopadhyay2020review}. Second, agents travel from their locations to the scene of incidents governed by a model of travel times. We assume that agents then take time to service the incident at some exogenously specified service rate. Finally, the system itself takes time to plan and implement the allocation of agents. We refrain from discussing the mathematical model and expressions for the temporal transitions and the state transition probabilities since our algorithmic framework only needs a generative model of the world (in the form of a black-box simulator) and not explicit estimates of transitions themselves.


\textbf{Rewards:}
Rewards in an SMDP can have two components: a lump sum immediate reward for taking actions and/or a continuous-time reward as the process evolves. The specific reward structure is highly domain-dependent. For example, taxi services might prioritize maximizing revenue, while paratransit services might focus on maximizing the total number of ride requests that can be served. The metric we are concerned with for ERM is \textit{incident response time} $t_r$, which is the time between the system becoming aware of an incident and when the first agent arrives on the scene. This is modeled using only an immediate reward, which we denote by $\rho(s, a)$, for taking action $a$ in state $s$:
\begin{equation}\label{eq:running action_util}
\rho(s,a) = \begin{dcases*}
\alpha^{t_h}(t_{r}(s,a)) &\text{if dispatching}\\
0 &\text{otherwise}
\end{dcases*}
\end{equation}
where $\alpha$ is the discount factor for future rewards, $t_h$ the time since the beginning of the planning horizon $t_{0}$, and $t_{r}(s,a)$ is the response time to the incident due to a dispatch action. The benefits of allocation actions are inferred from improved future dispatching.

\textbf{Problem Definition:}
 Given a state $s$ and a set of agents $\Lambda$, our goal is to find an action recommendation set $\sigma = \{a_1,...,a_m\}$ with $a_i \in A^i(s)$ that maximizes the expected reward. The $i$th entry in $\sigma$ contains a \textit{valid} action for the $i$th agent. In our ERM case study, this corresponds to finding an allocation of agents to depots that minimizes the expected response times to incidents.

%% file: Approach.tex
\section{Approach}\label{sec:approach}


\begin{figure}[t]
    \centering
    \includegraphics[width=\columnwidth]{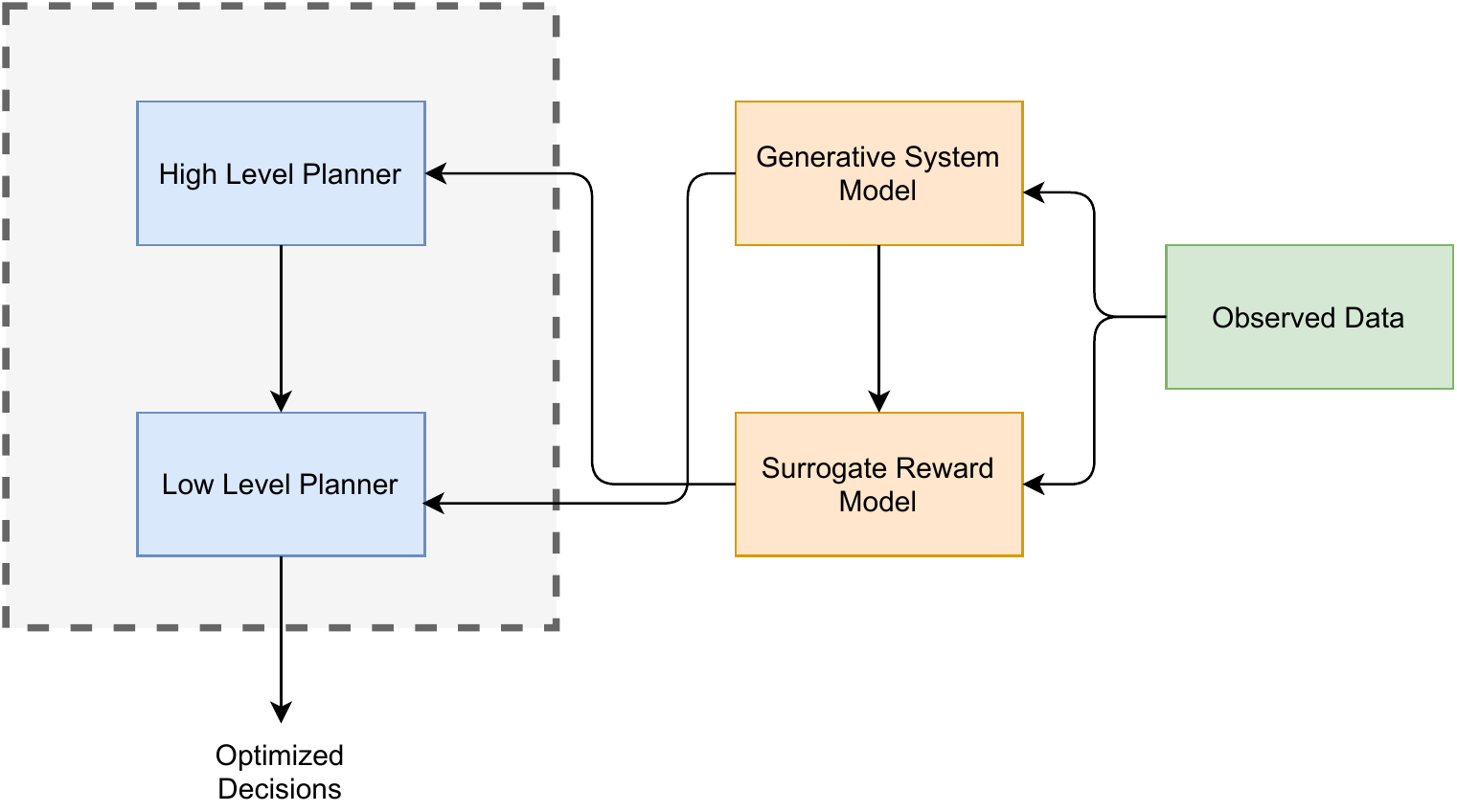}
    \caption{An overview of the proposed planning approach. We use the observed data to learn a generative model over when and where incidents occur. The generative model lets us simulate agent behavior, which aids the creation of a surrogate model for the high-level planner to segment the overall problem into a set of smaller problems. The low-level planner tackles each sub-problem independently. As we show, the surrogate reward model can also be estimated based on closed-form expressions of waiting times based on a queuing model.}
    \label{fig:hierarchical}
\end{figure}

Our approach to spatiotemporal resource allocation divides planning tasks into a two-stage hierarchy: ``high-level'' and ``low-level'' planning. First, high-level planning divides the overall decision-theoretic problem into smaller sub-problems. Our high-level planner accomplishes this by creating meaningful spatial clusters (which we call regions) and optimizing the distribution of agents among these regions. In order to do so, the high-level planner must assess the quality of a specific distribution of agents, which requires the low-level planner itself. Indeed, the actual reward generated from an allocation depends on how the low-level planner optimizes the distribution of responders \textit{within each region}. Clearly, this dependency is detrimental to creating an approach that seeks to divide the overall planning problem into two stages to achieve scalability. In order to tackle this challenge, we learn a surrogate model over waiting times that the high-level planner can use to estimate rewards. The high-level planner can perform inference using the model to estimate rewards as it optimizes the distribution of agents among the regions. Then, an instance of the low-level planner is instantiated for each of the regions. The low-level planner for a specific region optimizes the spatial locations of agents within that region. We assume the availability of an integrated simulation framework that models the dynamics of agents as they travel throughout the environment and a probabilistic generative model of incident occurrence learned from historical incident data. Our simulation framework and the incident model are described in section \ref{sec:framework}. 

Segmenting allocation into smaller sub-problems significantly reduces complexity compared to a centralized problem structure. Consider a city with $|\Lambda|=20$ agents and $|D|=30$ depots, each of which can hold one agent. With a centralized approach, any agent can go to any depot, so there are $\text{\textit{Permutations}}(|D|,|\Lambda|) = \frac{|D|!}{(|D|-|\Lambda|)!} = \frac{30!}{10!} = 7.31\times10^{25}$ possible assignments at each decision epoch. Now consider a hierarchical structure where the problem is split into $5$ evenly sized sub-problems, $|\Lambda_h|=4$ and $|D_h| = 6$ for each region. There are now $P(|D_h|,|\Lambda_h|) = 360$ possible allocations in each region, so there are $360\times5=1800$ possible actions across all regions. This reduction in complexity is about $22$ orders of magnitude compared to the centralized problem, at the cost of abstracting the interactions between agents in different regions through the high-level planner. Alternatively, a decentralized approach in which each responder plans only its actions reduces the complexity further to only $|D|=30$ possible allocations for each agent~\cite{pettet2020algorithmic}. However, this reduction in complexity comes at a cost as each agent must make assumptions regarding other agent behavior, leading to sub-optimal planning. Hierarchical planning offers a balance between decentralized and centralized planning. 

An important consideration when designing approaches for resource allocation under uncertainty in city-scale CPS problems is adapting to the dynamic environments in which such systems evolve. In our decision support system, a decision coordinator (an automated module) invokes the high-level planner at all states that allow the scope for making decisions. For example, consider that an agent is unavailable due to maintenance. The coordinator triggers the high-level planner and notifies it of the change. The high-level planner then checks if the spatial distribution of the agents can be optimized to best adapt to the situation at hand. We describe the exact optimization functions, metrics, and approaches that we use to design the planners below.

\input{high_level}
\input{low_level}

%% file: high_level.tex
\subsection{High-Level Planner}

We seek to decompose the overall MSMDP into a set of tractable sub-problems through the high-level planner. A natural decomposition for spatiotemporal resource allocation is to divide the overall problem's spatial area into discrete regions and allocate separately for each region. This decomposition allows the low-level allocation planner to evaluate potential interactions between nearby agents and, therefore, likely to interact. The first goal of the high-level planner is to define these spatial regions. Consider that the high-level planner seeks to divide the overall problem in to $m$ regions, denoted by the set $R = \{r_1,r_2,\dots , r_m\}$, where $r_j \in R$ denotes the $j$th region. 
To achieve this goal, we use the locations of historical incident data.
Intuitively, we want to create regions based on \textit{hotspots} of incident occurrence~\cite{chainey2002hotspot}. Recall that our goal is to identify spatial areas where planning (the allocation and distribution of agents) can be performed independently. 
By identifying spatial incident clusters, we achieve the following: 1) areas close to each other with similar patterns of incident occurrence are grouped together, and 2) areas of high incident occurrence that are further apart get segregated from each other. While any standard spatial clustering approach can be used to achieve this goal, we use the well-known $k$-means algorithm in our analysis~\cite{lloyd1982least}. The $k$-means algorithm partitions a given set of points in $\mathbb{R}^d$ into $k$ clusters such that each point belongs to its closest cluster (defined by distance to the center of the cluster). While the problem is known to be NP-hard even when $d=2$ (our case)~\cite{mahajan2009planar}, heuristic-based iterative approaches can be used to achieve good solutions. Typically, the solution process repeatedly performs two steps after initializing an initial set of clusters. In the first step, each incident from the training set is assigned to the cluster whose center is the closest to the incident's location by Euclidean distance. Then, given an assignment, the centers of the clusters are computed again. The process is repeated until convergence. Clusters are then mapped to the cells $G$. Each cell in $G$ is assigned to the cluster containing the most incidents that occurred within the cell. We use these clusters of cells in $G$ as separate sub-problems for low-level planning.

The high-level planner's second task is to determine the distribution of agents across the decomposed spatial regions. If the regions are homogeneous, agents could be split evenly across them. In practice, however, regions will differ with respect to properties such as size, incident rate, and depot distributions, all of which impact the number of responders needed to cover each region. For example, the number of agents assigned to cover a dense downtown area should likely be different from sparsely populated suburbs. Recall that the overall goal of the parent MSMDP is to reduce the expected incident response times. Response times to emergency incidents consist of two parts: a) the time taken by an agent to travel to the scene of the incident, and b) the time taken to service the incident. Suppose we assume that incidents are homogeneous, meaning that the time taken by agents to service incidents follows the same distribution; in that case, the sole criterion that a planner needs to optimize is the travel time of the agents to incidents, which we refer to as \textit{waiting times} (achieving zero waiting times is clearly infeasible in practice, so we seek to minimize waiting times). Therefore, the high-level planner seeks to distribute agents to different regions to minimize the expected incident waiting time. It is important to note that the real world is more complex; features such as incident severity can impact service times as well as the priority of responding to each incident. Incorprating such features and modeling service time variations in the high-level planner is a topic for future work. 

We denote the expected waiting time for incidents in region $r_j \in R$ by $w_{j}(p_j, \gamma_j)$, where $p_j$ is the number of responders assigned to the region $r_j$ and $\gamma_j$ is the total incident rate across $r_j$. Since the arrival process is assumed to be Poisson distributed, $\gamma_j$ can be calculated as $\sum_{g_i \in G} \mathbb{1}(g_i \in r_j)\gamma_i$, where $\mathbb{1}(g_i \in r_j)$ denotes an indicator function that checks if cell $g_i$ belongs to region $r_j \in R$. We consider two approaches to estimate $w_{j}(p_j, \gamma_j)$: a \textit{queuing model} that approximates the system using an \textit{m/m/c} queuing formulation, and a \textit{surrogate model} that uses machine learning. We detail these models in sections \ref{sec:queue} and \ref{sec:surrogate}, and then describe our high-level agent distribution algorithm, which uses an iterative greedy approach to assign agents across regions, in section \ref{sec:hl_optimization}.

\subsubsection{Queuing Model}\label{sec:queue}
A multi-server queue model is one approach to model waiting times in a region. Recall that incident arrivals are distributed according to a Poisson distribution, thereby making inter-incident times exponentially distributed. We make the standard assumption that service times are exponentially distributed as well~\cite{mukhopadhyayAAMAS17}. One potential issue with using well-known queuing models to estimate waiting times in emergency response is that travel times are not memoryless. In this approach, we use an approximation from prior work to tackle this problem~\cite{mukhopadhyayAAMAS17}. Specifically, travel times to emergency incidents are typically much shorter than service times. Thus, the sum of travel times and service times can be considered to be approximated by a memoryless distribution (provided that the service time itself can be modeled by a memoryless distribution). The average waiting time for a region $r_j \in R$ can then be estimated by considering a $m/m/c$ queuing model (using Kendall's notation~\cite{kendall1953stochastic}), where $c=p_j$. 


Let the average service time be $T_s$ and let $\mu=1/T_s$ denote the mean service rate. Then, the mean waiting time $w_j(p_j, \gamma_j)$ is ~\cite{shortle2018fundamentals}: 
\[
    w_j(p_j, \gamma_j) = \frac{P_0 (\frac{\gamma_j}{\mu})^{p_j}\gamma_j}{c!(1-\rho)^2 c}
\]
where $P_0$ denotes the probability that there are 0 incidents waiting for service and can be represented as
\[
    P_0 = 1/\Big[\sum_{m=0}^{p_j -1} \frac{(p_j \rho)^m}{m!} + \frac{(c \rho)^c}{c!(1 - \rho)} \Big]
\]

While this approach is straightforward, computationally efficient, and works out of the box with any city of interest, it abstracts away many environmental dynamics that can affect response times. For example, the agents' allocation within a region, travel times due to traffic, and behaviors such as dropping off patients at the hospital can all affect the response times observed in the real world. To capture such factors, we explore an alternative method using machine learning to create a surrogate model over expected waiting times. 

\subsubsection{Surrogate Model}\label{sec:surrogate}
Another approach for estimating waiting times in a region is to use simulated data to learn a model of waiting times conditioned on a set of relevant covariates. For example, we simulate emergency response in a region under various conditions (namely, the number of agents assigned to the region and the incident rate in the region) and record the simulated incident response times. After generating many such samples, we use a supervised learning approach to fit a parameterized function to maximize the likelihood of the simulated data. The trained model can then be used to estimate waiting times for a region, given the number of responders. An advantage of this approach is that it captures subtle environmental dynamics that are ignored by the queue model, such as the travel times of agents as they respond to incidents and their location within the region due to factors such as depot locations and dropping off patients at hospitals. It is also adaptable to changes in the system model (e.g. extending the model to consider severity or heterogeneous service times), as these can be incorporated into the underlying simulation used to train the surrogate model. 

The first step in creating the surrogate model is generating response time samples by simulating emergency response in each region with different incident rates $\gamma_j$ and available agents. We sample incidents using the model described in section \ref{sec:framework}, which uses historical rates of incidents to create a sampling distribution. We refer to sampled incidents as \textit{chains}. For each such chain, region $r_j$, and potential number of agents 
$p_j \in \{1,\dots,|D_j|\}$
(where $D_j$ is the set of depots in region $r_j$), we simulate response to the incidents occurring within $r_j$ using the simulation framework described in section \ref{sec:framework}. The simulation provides us with labeled training data where each observation captures the response time (output) given the number of responders and incident rate (inputs).

A crucial factor that affects simulated response times is the initial location of the agents within each region. Recall that in our proposed framework, the location of the agents is optimized by the low-level planner, which is naturally infeasible to use for every step during sampling. As a result, we solve the standard $p$-median problem~\cite{dzator_effective_2013}, which is often applied to ambulance allocation, to determine the initial conditions of the simulation. The objective of the $p$-median formulation is to locate $p_j$ agents (in the region $r_j$) such that the average demand-weighted distance between demand points and their nearest agent is minimized. Formally, we solve the following optimization problem to allocate the agents to depots:
\begin{subequations}
\begin{align}
    \text{min} & \sum_{i=1}^{|G_j|} \sum_{k=1}^{|D_j|} a_{i} d_{ik} Y_{ik} \label{allocation_objective} \\
    \text{s.t.} &  \sum_{k=1}^{|D_j|} Y_{ik} = 1, \ \ \ \forall i \in \{1,\dots,|G_j|\} \label{allocation_meet_demand} \\
    & \sum_{k=1}^{|D_j|} X_k = p \label{allocation_p} \\
    & Y_{ik} \leq X_k, \ \ \ \forall i \in \{1,\dots,|G_j|\}, \forall k \in \{1,\dots,|D_j|\} \label{allocation_placed_responder} \\
    & X_{k}, Y_{ik} \in \{0, 1\}, \ \ \ \forall i \in \{1,\dots,|G_j|\}, \forall j \in \{1,\dots,|D_j|\}
\end{align}
\end{subequations}
\normalsize
where $G_j$ is the set of cells in region $r_j$, $D_j$ is the set of depots in $r_j$, $p_j$ is the number of agents to be located in $r_j$, $a_{i}$ is the likelihood of accident occurrence at cell $g_i \in G_j$, and $d_{ik}$ is the distance between cell $g_i \in G_j$ and location $d_k \in D_j$. $Y_{ik}$ and $X_k$ are two sets of decision variables; $X_k = 1$ if an agent is located at $d_k \in D_j$ and $0$ otherwise, and $Y_{ik} = 1$ if cell $g_i \in G_j$ is covered by an agent located at $d_k \in D_j$ (i.e. the agent at $d_k$ is the nearest placed agent to $g_i$) and $0$ otherwise.

\small
\begin{algorithm}[]
\caption{Greedy-Add Algorithm}
\label{algo:greedy_add}
\SetAlgoLined
\KwIn{Region cells $G_j$, region depots $D_j$, cell incident likelihoods $a_i \ \ \forall i \in \{1,\dots,|G_j|\}$, cell to depot distances $d(i, k) \ \ \forall i \in \{1,\dots,|G_j|\}$ and $k \in \{1,\dots,|D_j|\}$, number of agents $p_j$}
\KwOut{Agent depot assignments $X$}
Initialize $z := 0$, $X_z := \emptyset$ \; \label{algo:ga_init}
\While{$z < p_j$}{
    $z := z + 1$\; \label{algo:ga_iter_update}
    
    \For{$\mathrm{location } \ d_{k'} \in D_j, \ \ \mathrm{where } \ k' \notin X_{z-1}$}{
        $X'_{z} := X_{z-1} \cup d_{k'}$\; \label{algo:ga_start_score}
        Find nearest facilities $y_i \ \forall i \in \{1,\dots,|G_j|\}$, where $y_{i} \in X'_z$\;
        Compute $U^z_{k'} := \sum_{g_i \in G_j} a_i d(g_i, y_i)$\; \label{algo:ga_end_score}
    
    }
    Best location $d_k^* := \text{argmin}_k \ U^z_k$\; \label{algo:ga_find_best}
    $X_z := X_{z-1} \cup k^*$\; \label{algo:ga_add_best}
Return $X_z$
}
\end{algorithm}
\normalsize

The p-median problem is NP-hard~\cite{kariv1979algorithmic}; therefore, heuristic methods are employed to find approximate solutions in practice. We use the Greedy-Add algorithm~\cite{daskin2011network} to optimize the locations of agents. We show the algorithm in Algorithm \ref{algo:greedy_add}. First, we initialize the iteration counter $z$ and the set of allocated agent locations $X_z$ to the empty set (step \ref{algo:ga_init}). Then, as long as there are responders awaiting allocation, we iterate through the following loop: (1) update counter $z$ to the current iteration (step \ref{algo:ga_iter_update}), (2) for each potential location not already in the allocation, compute the p-median score (equation \ref{allocation_objective}) of the allocation which includes the potential location (steps \ref{algo:ga_start_score} - \ref{algo:ga_end_score}), and (3) find the location that minimizes the p-median score (step \ref{algo:ga_find_best}) and add it to the set of allocated agent locations (step \ref{algo:ga_add_best}). After locating the agents to depots within the region, we simulate emergency response using greedy dispatch and record the average response times $w(r_j, p_j, \lambda_j)$. The complexity of this algorithm is $O(p_j |D_j| |G_j|)$ because assigning each agent requires evaluating each potential depot in $D_j$, which requires summing over the weighted distances between each cell in $G_j$ and its nearest populated depot.

Given a set of samples of waiting times ($w$), agent allocations ($p$), and incident rates ($\gamma$), the second step in creating the surrogate model is to learn an estimator over $w$ given $p$ and $\gamma$. We learn a different model for each region to capture any latent features that can effect response times such as the region's depot distribution and roadway network. Specifically, we use random forest regression~\cite{breiman2001random}, which is an ensemble learning method based on constructing several decision trees at training time. During inference, the average prediction of the trained trees is used as the output (for regression problems).

\subsubsection{Optimization}\label{sec:hl_optimization}

\setlength{\textfloatsep}{5pt}
\begin{algorithm}[t]
\SetAlgoLined
\KwIn{Sorted regions $R_s$, arrival rates $\{\gamma_1,\gamma_2,\dots,\gamma_m\}$, service rate $\eta$}
\KwOut{Responder allocation $P\,=\,\{p_1,p_2,\dots,p_m\}$}
 $\text{assigned} := 0, i := 0, J := \emptyset$\;
 \While{\text{assigned} $\leq \,\,\mid \Lambda \mid \,\, \textbf{and} \,\, i \leq m$}{
  $p_i := p_i + 1$\;\label{algo:high_assign}
  $\text{assigned} := \text{assigned} + 1$\;
  \If{$\eta \times (p_i) \geq \sum_{g_i \in G} \mathbb{1}(g_i \in r)\gamma_i$ }{\label{algo:high_sustain}
  $i := i + 1$\;\label{algo:high_next}
  }
 }
 \While{\text{assigned} $\leq \,\,\mid \Lambda \mid$}{\label{algo:high_surplus}
  \For{$i \in \{1,\dots, m\}$}{$J[i] := w_i(p_i, \gamma_i) - w_i(p_i + 1, \gamma_i)$\;}
  $r^{*} := \argmax_{i \in \{1,\dots, m\}}J[i]$\;\label{algo:high_marginal}
  $p_{r^{*}} := p_{r^{*}} + 1$\;
  $\text{assigned} := \text{assigned} + 1$\;
 }
 \caption{High-Level Planner}
 \label{algo:high}
\end{algorithm}

Given estimated waiting times for incidents in each region, the high-level planner seeks to minimize the cumulative response times across all regions. The optimization problem can be represented as:
\begin{subequations}
\label{eq:highLevel}
\begin{align}
	&\min_{p}  \sum_{j=1}^{m} w_j(p_j)\\
	&\text{s.t.}\,\,\,\,\sum_{i=1}^{m} p_i = |\Lambda| \label{cons:budget}\\
	& p_i \in \mathbb{Z}^{0+} \,\, \forall i \in \{1,\dots,m\}
\end{align}	
\end{subequations}
The objective function in mathematical program \ref{eq:highLevel} is non-linear and non-convex. We use an iterative greedy approach shown in algorithm \ref{algo:high}. We begin by sorting regions according to total arrival rates. Let this sorted list be $R_s$. Then, we assign agents iteratively to regions in order of decreasing arrival rates (step \ref{algo:high_assign}). After assigning each agent to a region $r_j \in R$, we compare the overall service rate ($p_j$ times the mean service rate by one agent) and the incident arrival rate for the region (step \ref{algo:high_sustain}). Essentially, we try to ensure that given a pre-specified service rate, the expected length of the queue is not arbitrarily large. Once a region is assigned enough responders to sustain the arrival of incidents, we move on to the next region in the sorted list $R_s$ (step \ref{algo:high_next}). Once all regions are assigned agents in this manner, we check if there are surplus agents (step \ref{algo:high_surplus}). The surplus agents are assigned iteratively according to the incremental benefit of each assignment. Specifically, for each region, we calculate the marginal benefit $J$ of adding one agent to the existing allocation (step \ref{algo:high_marginal}). Then, we assign an agent to the region that gains the most (in terms of reducing waiting times) by the assignment. 

The complexity of the algorithm is $O(|\Lambda|m\xi)$, where $\xi$ is the complexity of the wait time estimation method, as the algorithm computes the potential wait times $w_j(r_j, p_j, \lambda_j)$ from adding an agent to each region when assigning said agent. The overall complexity is $O(|\Lambda|^2 m)$ using the queuing model, since equation $P_0$ sums over all assigned agents to estimate $w_j$.  When using the random forest surrogate model, the overall complexity is $O(|\Lambda|m \epsilon \beta )$, where $\epsilon$ is the number of trees in the forest and $\beta$ is the maximum tree depth. An important note is that in our problem instances, we assume that there are enough agents to process the incident demand generated in $G$. The proposed greedy approach may leave some regions with no responders in environments where this assumption does not hold. Future work can examine if the proposed approach or an alternative (such as proportional assignment with respect to incident rate) performs better in such situations.


%% file: low_level.tex
\subsection{Low-Level Planner}\label{sec:ll}

The fine-grained allocation of agents to depots within each region is managed by the low-level planner, which induces a decision process for each region that is smaller than the original MSMDP problem described in section \ref{sec:problem_formulation} by design. The MSMDP induced by each region $r_j \in R$ contains only state information and actions that are relevant to $r_j$, i.e., the depots within $r_j$, the agent's assigned to $r_j$ by the inter-region planner, and incident demand generated within $r_j$. Decomposing the overall problem makes each region's MSMDP tractable using many approaches, such as dynamic programming, reinforcement learning (RL), and Monte Carlo Tree Search (MCTS). Each approach has advantages and trade-offs that must be examined to determine which is best suited for the specific problem domain under consideration. 

Spatial-temporal resource allocation has a key property that informs the choice of the solution method---a highly dynamic environment that is difficult to model in closed-form. To illustrate, consider a travel model for the agents. While there are long-term trends for travel times, precise predictions are difficult due to the complex interactions between features such as traffic, weather, and events occurring in the city. Furthermore, a city's traffic distribution changes with changes in the road network and population shifts, so it needs to be updated periodically with new data. This dynamism is true for many pieces of the domain's environment, including the demand distribution of incidents. Importantly, it is also true of the system itself: agents can enter and leave the system due to mechanical issues or purchasing decisions, and depots can be closed or opened. Whenever underlying environmental models change, the solution approach must consider the updates to make correct recommendations. Approaches that require long training periods, such as reinforcement learning and value iteration, are challenging to apply to such scenarios since they must be re-trained each time the environment changes. This challenge motivates using Monte Carlo Tree Search (MCTS), a probabilistic search algorithm, as our solution approach. Being an anytime algorithm, MCTS can immediately incorporate any changes in the underlying generative environmental models when making decisions. 

MCTS represents the planning problem as a ``game tree'', where nodes in the tree represent states. The decision-maker is given a state of the world and is tasked with finding a promising action. The current state is treated as the root node, and actions that are taken from one state to another are represented as edges between corresponding nodes. The core idea behind MCTS is that the tree can be explored asymmetrically, with the search being biased toward actions that appear promising. To estimate the value of an action at a state node, MCTS simulates a ``playout'' from that node to the end of the planning horizon using a computationally inexpensive \textit{default policy} (our simulated system model is shown in figure \ref{fig:state} and described in detail in section \ref{sec:framework}). This policy is not required to be very accurate (indeed, a standard method is random action selection), but as the tree is explored and nodes are revisited, the estimates are re-evaluated and converge toward the actual value of the node. This asymmetric tree exploration allows MCTS to search  large action spaces quickly. 
When implementing MCTS, there are a few domain specific decisions to make---the \textit{tree policy} used to navigate the search tree and find promising nodes to expand, and the \textit{default policy} used to quickly simulate playouts and estimate the value of a node. 

\textbf{Tree Policy:} When navigating the search tree to determine which nodes to expand, we use the standard Upper Confidence bound for Trees (UCT) algorithm \cite{kocsis2006bandit}, which defines the score of a node $n$ as 
\begin{equation}
    \text{UCB}(n) = \overline{u}(n) + c\sqrt{\frac{\text{log}(\text{visits}(n))}{\text{visits}(n')}}
\end{equation}
where $\overline{u}(n)$ is the estimated utility of state at node $n$, visits($n$) is the number of times $n$ has been visited, and $n'$ is node $n$'s parent node. When deciding which node to explore in the tree, the child node with the maximum UCB score is chosen. The term $\overline{u}(n)$ corresponds to the exploitation objective that favors nodes that have produced higher rewards in the past. The second term on the right-hand side of the equation corresponds to the exploration objective, encouraging the exploration of nodes with low visit counts. The constant $c$ controls the tradeoff between these two opposing objectives and is domain-dependent. 

\textbf{Default Policy:} When working outside the MCTS tree to estimate the value of an action, i.e., rolling out a state, a fast heuristic \textit{default policy} is used to estimate the score of a given action. Rather than using a random action selection policy, we exploit our prior knowledge that agents generally stay at their current depot unless significant shifts in incident distributions occur. Therefore, we use greedy dispatch without redistribution of responders as our heuristic default policy.

\begin{algorithm}[t]
\SetAlgoLined

\KwIn{Regions $R$, state $s$, generative demand model $E$, number of samples $n$}
\KwOut{Recommended allocation actions $\sigma_r \,\, \forall r \in R$}
 
 \For{$\text{region } r_j \in R$}{
    $\text{Decompose } s \text{ into region specific state } s_j$\;\label{algo:low_decompose}
    $\text{Action score map } \widetilde{\mathcal{A}} := \emptyset$\;
    $\text{eventChains } := E\text{.sample}(s_j, n)$\;\label{algo:low_sampleChains}
    $\text{Action scores } A := \text{MCTS}(s_j, \text{eventChains})$\;
    \For{$\text{action }a \in A$}{
        $\widetilde{\mathcal{A}}[a]\text{.append}(\text{score}(a))$\;
    }
    $\overline{\mathcal{A}} := \emptyset$\;
    \For{$\text{potential action }a \in A$}{
        $\overline{\mathcal{A}}[a] = \text{mean}(\widetilde{\mathcal{A}}[a])$\;
    }
    
    $\text{Recommended action }\sigma_r := \text{argmax}_{a} \ \ \overline{\mathcal{A}}[a]$\label{algo:low_argmax}

 }

 \caption{Low-Level Planner}
 \label{algo:low}
\end{algorithm}

It is important to note that performing MCTS on one sampled chain of events is insufficient in practice, as traffic incidents are inherently sparse. Any particular sample is too noisy to determine the value of an action accurately. To handle this uncertainty, we use \textit{root parallelization}. We sample many incident chains from the prediction model and instantiate separate MCTS trees to process each. We then average the scores across trees for each potential allocation action to determine the optimal one. Our low-level planning approach is shown in algorithm \ref{algo:low}. The inputs for low-level planning are the regions $R$, the current overall system state $s$ (which includes each agent's region assignment), a generative demand model $E$, and the number of chains to sample and average over for each region $n$. For each region $r_j \in R$, we first extract the state $s_j$ in the region's MSMDP from the current overall system state $s$ (step ~\ref{algo:low_decompose}). Then we perform root parallelization by sampling $n$ incident chains from the demand model $E$ and performing MCTS on each to score each potential allocation action (step ~\ref{algo:low_sampleChains}). It is important to note that the sampled incident chains are specific to the region under investigation, and demand is only generated from the cells in that region. We then average the scores across samples for each action and choose the allocation action with the maximum average score (step ~\ref{algo:low_argmax}).

%% file: Implementation.tex
\section{Integration Framework}\label{sec:framework}


\begin{figure}[t]
    \centering
    \includegraphics[width=\columnwidth]{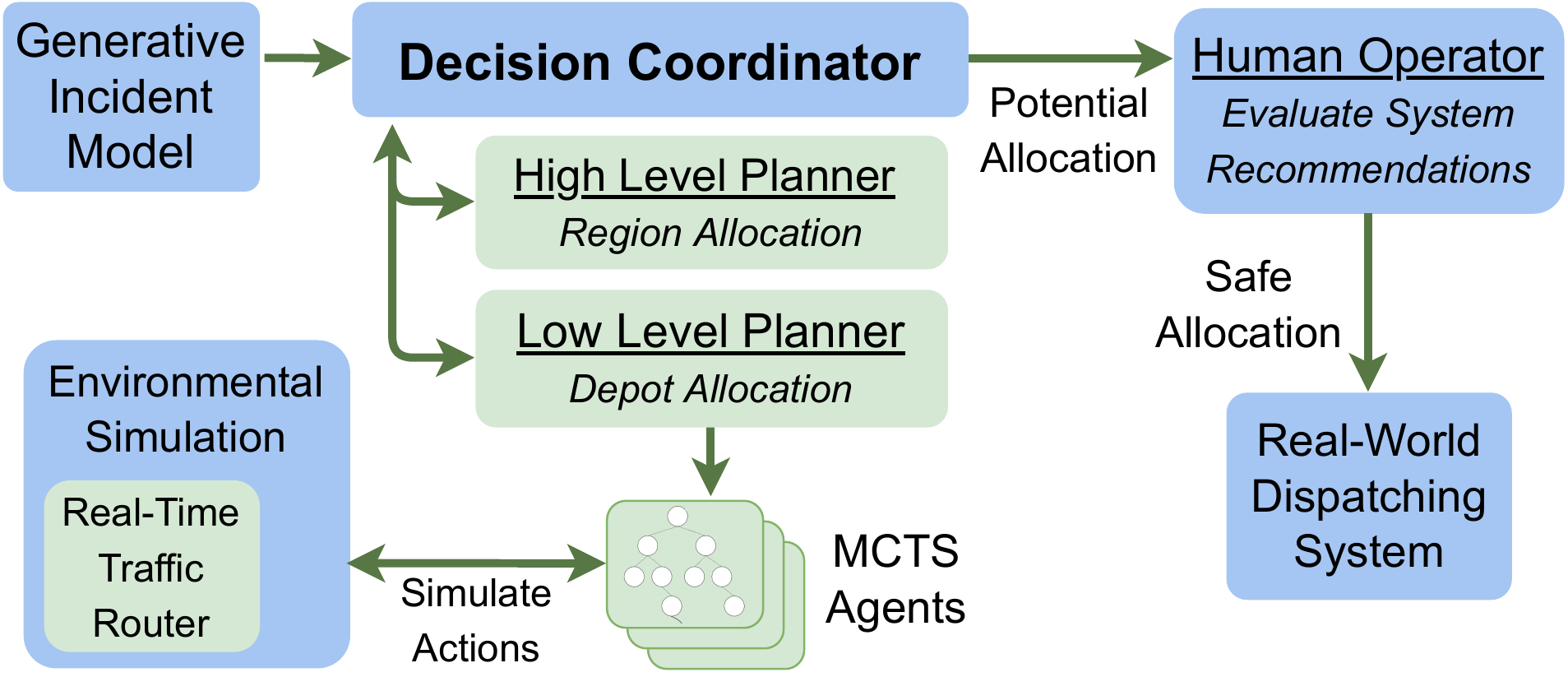}
    \caption{Emergency response decision support framework.}
    \label{fig:software_diagram}
\end{figure}


Figure \ref{fig:software_diagram} shows a schematic representation of our decision support system. Realizing a such as system for online emergency responder allocation requires a framework of interconnected processes, including:

\begin{itemize}
    \item A traffic routing model to support routing requests (section \ref{sec:travel}).
    \item A probabilistic generative model of incident occurrence (section \ref{sec:forecast}).
    \item A model of the ERM system and its environment, including the dynamics of responders, depot locations, and hospital locations (section \ref{sec:system_model}). 
    \item A simulation of the ERM system built on the above components (section \ref{sec:simulation}).
    \item A hierarchical decision process that makes allocation and dispatching recommendations based on the current state of the environment, responder locations, and projected incident distributions (section \ref{sec:approach}).
    \item A human operator to access the planning mechanism and act as an interfaced with a real-world computer-aided dispatch system. 
\end{itemize}
\subsection{Travel Model} \label{sec:travel}

We consider two travel time models in our implementation. The first model is based on the Euclidean distance between two points of interest (the centers of the cells in the grid). We also develop a more principled travel model that uses contraction hierarchies~\cite{geisberger2008contraction} and an open-source routing machine engine~\cite{huber2016calculate} to look up travel times at different times of the day from the center of each cell in the spatial grid to other cells. We collect such travel times across a week. The final travel model considers the median of the accumulated travel times and develops a lookup table which the overall planning process can use to query the travel times. This approach is better than a Euclidean distance-based travel time as it considers the average road congestion and the maximum travel speed across the road. The pipeline we propose and our framework are flexible to accommodate other travel time models; for example, a modular component that estimates travel times based on advanced graphical neural networks can be used with a richer set of covariates to provide sensitive estimates of time and weather.


\subsection{Incident Prediction}\label{sec:forecast}

Recall that we need samples of incidents for the low-level planner and learning the surrogate model for the high-level planner. We assume that incidents are generated by a Poisson distribution. A Poisson model has been widely used to model the occurrence of accidents~\cite{mukhopadhyay2020review}. The Poisson distribution is a discrete probability distribution over the number of events in a fixed time interval. We learn a separate Poisson model for each cell $g_i \in G$. The rate parameter of each model can be learned by maximizing the likelihood of historical data in the cell. The learned Poisson model can then be used to sample incidents in a given period of time.

\subsection{ERM System Model}\label{sec:system_model}


\begin{figure}[t]
    \centering
    \includegraphics[width=\columnwidth]{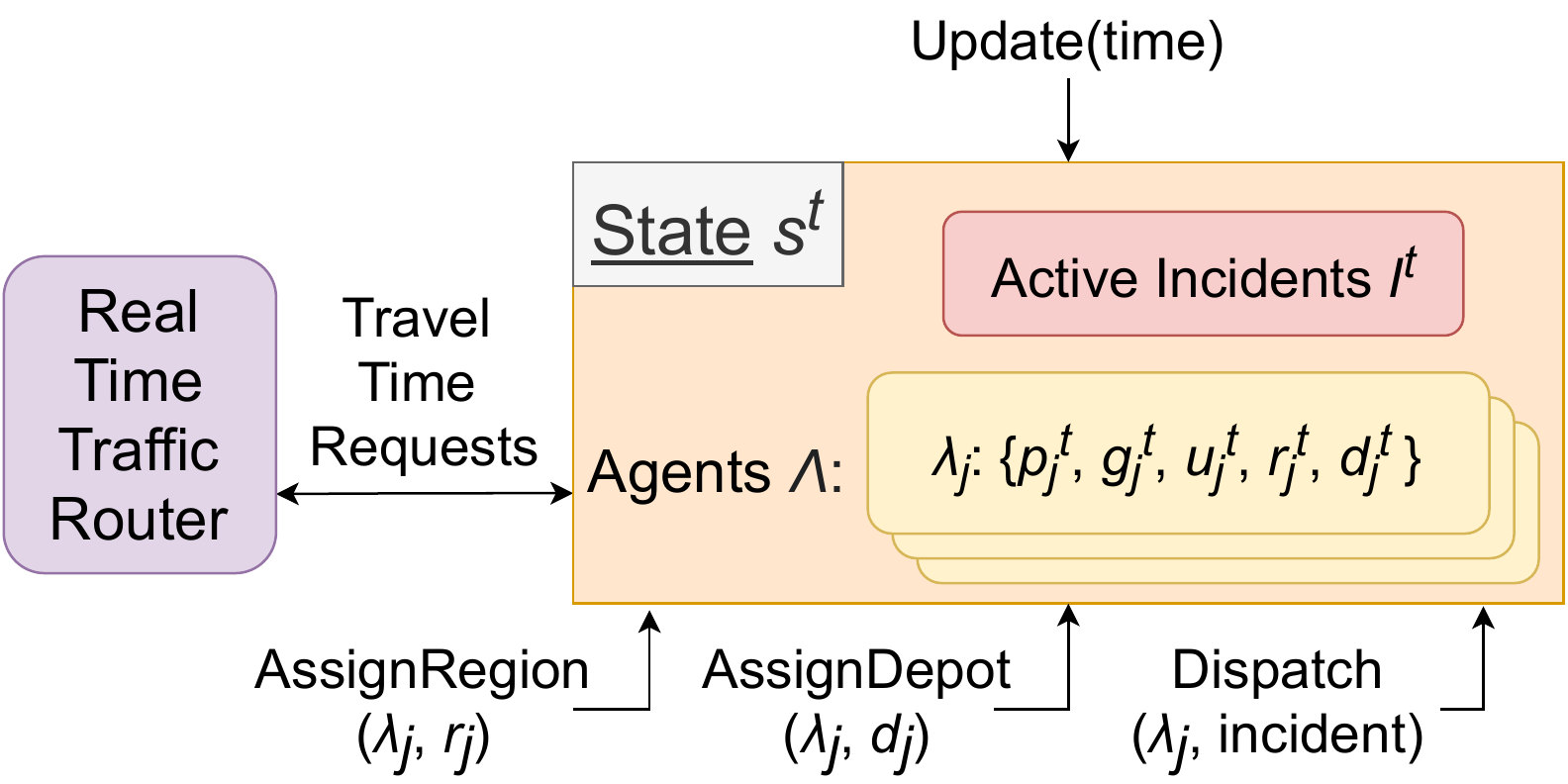}
    \caption{System state and actions.}
    \label{fig:state}
\end{figure}

As shown in figure \ref{fig:state}, our system state at time $t$ is captured by a queue of active incidents $I^{t}$ and agent states $\Lambda$. $I^t$ is the queue of incidents that have been reported but not yet serviced. The state of each agent $\lambda_j \in \Lambda$ consists of the agent's current location $p_j^t$, status $u_j^t$, destination $g_j^t$, assigned region $r_j^t$, and assigned depot $d_j^t$. Each agent can be in several different internal states (represented by $u_j^t$), including \textit{waiting} (waiting at a depot), \textit{in\_transit} (moving to a new depot and not in emergency response mode), \textit{responding} (the agent has been dispatched to an incident and is moving to its location), and \textit{servicing} (the agent is currently servicing an incident). These states dictate how the agent is updated when moving the simulator forward in time. 

\subsection{Simulation Framework} \label{sec:simulation}

Our simulator is designed as a discrete event simulator, meaning that the state is only updated at discrete time steps when certain events occur. These events include incident occurrence, re-allocation planning steps, and responders becoming available for dispatch. Between these events, the system evolves based on prescribed rules. Using a discrete event simulator saves valuable computation time compared to a continuous-time simulator. At each time step when the simulator is invoked, the system's state is updated to the current time. First, if the current event of interest is an incident occurrence, it is added to the active incidents queue $I^t$. Then, each agent's state and locations are updated accordingly. For example, agents that are in the \textit{waiting} state stay at the same position, while agents that are \textit{responding} or \textit{in\_transit} are updated according to the travel time model. If they reach their intended destinations, their states are updated to \textit{servicing} or \textit{waiting} at their respective locations. If such responders have not reached their destination at the time under consideration, their locations are updated using the travel model. If an agent is in the \textit{servicing} state and finishes servicing an incident, its state is updated to the \textit{in\_transit} state, and its destination $g_j^t$ is set to the assigned depot.  

After the state is updated, a planner has several actuations available to control the system. The \textit{Dispatch}$(\lambda_j, \textit{incident})$ function will dispatch the agent $\lambda_j$ to the given incident which is in $I^t$. Assuming the responder is available, the system sets agent $\lambda_j$'s destination $g_j^t$ to the incident's location, and its status $u_j^t$ is set to \textit{responding}. The incident is also removed from queue $I^t$ since it is being serviced, and the response time is returned to the planner for evaluation. The planner can also change the allocation of the agents. \textit{AssignRegion}$(\lambda_j, r_j)$ assigns agent $\lambda_j$ to region $r_j$ by updating $\lambda_j$'s $r_j^t$. \textit{AssignDepot}$(\lambda_j, d_j)$ similarly assigns agent $\lambda_j$ to depot $d_j$ by updating $\lambda_j$'s $d_j^t$ and setting its destination $g_j^t$ to the depots location. These functions allow a planner to try different allocations and simulate various dispatching decisions. 

%% file: Experiments.tex
\section{Experiments} \label{sec:exp}

\begin{figure*}
    \centering
    \includegraphics[width=\textwidth]{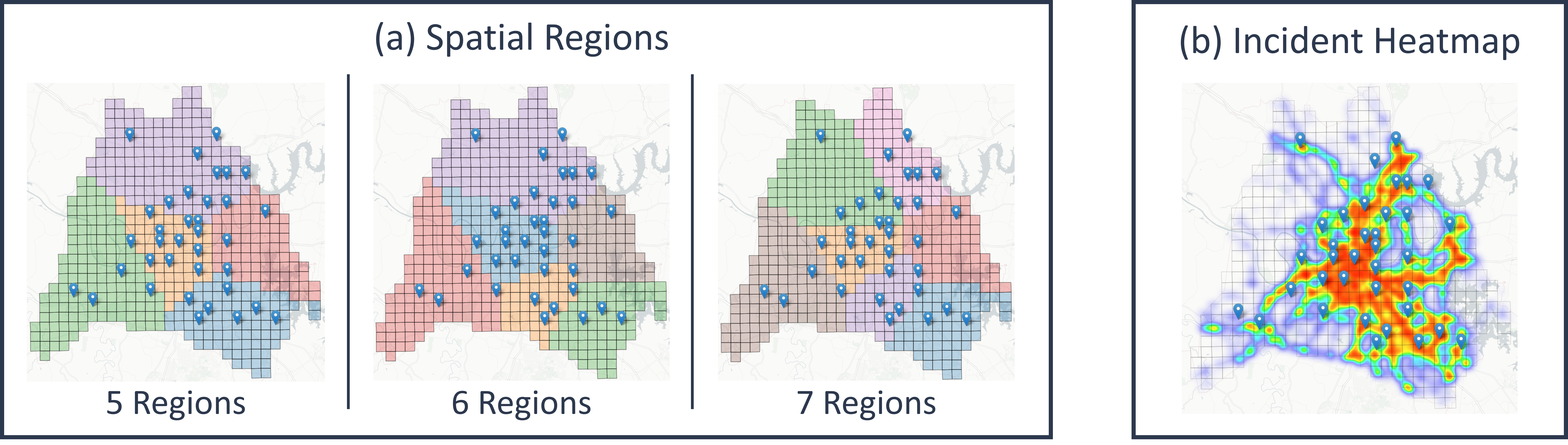}
    \caption{Subfigure (a) -- The various spatial regions under consideration. Pins on the map represent depot locations, and different colors represent different spatial regions. Subfigure (b) -- Nashville's historic incident density from January 2018 to May 2019 overlaid on the spatial grid environment.}
    \label{fig:combined_heatmap_regions}
\end{figure*}

We evaluate the proposed hierarchical framework's effectiveness on emergency response data obtained from Nashville, Tennessee, a major metropolitan area in the United States, with a population of approximately 700,000. We use historical incident data, depot locations, and operational data provided by the Nashville Fire Department~\cite{fireDepartmentCommunication}. We construct a grid representation of the city using $1\times1$ mile square cells; this choice resulted from the fact that local authorities follow a similar granularity of discretization. These cells, as well as the city's 35 depot locations, can be seen in figure~\ref{fig:combined_heatmap_regions}.

\textbf{Configuration and hyper-parameters:} We make a few important assumptions when configuring our experiments. First, we limit the capacity of each depot $C(d)$ to 1. This constraint encourages responders to be geographically spread out to respond quickly to incidents occurring in any region of the city, and it models the usage of ad-hoc stations by responders, which are often temporary parking spots.\footnote{In theory, we could always add dummy depots at the same location to extend our approach to a situation where more than one responder per depot is needed.} We assume there are 26 available responders to allocate, which is the actual number of responders in the urban area under consideration~\cite{fireDepartmentCommunication}. Third, we assume that the mean rate to service an incident is 20 minutes based on actual service times in practice in Nashville (we hold this constant in our experiments to compare the planning approaches directly). Fourth, as mentioned in section~\ref{sec:approach}, we assume that incidents are homogeneous. 

The number of MCTS iterations performed when evaluating potential actions on a sampled incident chain is set to 1000, and the number of samples from the incident model that are averaged together using root parallelization during each decision step is set to 50. We run the hierarchical planner after each test incident to re-allocate responders. Further, if the planner is not called after a pre-configured time interval, we call it to ensure that the allocations are not stale. In our experiments, this maximum time between allocations is set to 60 minutes. We ran experiments on an Intel i9-9980XE, with 38 logical processors running at a base clock of 3.00 GHz and 64 GB RAM.
Our experimental hyper-parameter choices are shown in table ~
\ref{tab:hyperparams}. In our experiments, we vary the number of spatial regions to examine how their size and distribution effects performance of the hierarchical planner; the resulting region configurations can be seen in figure \ref{fig:combined_heatmap_regions}. 

\textbf{Incident Model:} Our incident model is learned from historical incident data. We learn a Poisson distribution over incident arrival for each cell based on historical data. The maximum likelihood estimate (MLE) of the rate of the Poisson distribution is simply the empirical mean of the number of incidents in each unit of time. To simulate our system model, we access the Poisson distribution of each cell and sample incidents from it. In reality, emergency incidents might not be independently and identically distributed; however, the incident arrival model (and the black box simulator of the system in general) is entirely exogenous to our model and does not affect the functioning of our approach. To validate the robustness of our approach, we create three separate testbeds based on domain knowledge and preliminary data analysis of historical incident data.

\textbf{Region Segmentation:} We use the $k$-means algorithm~\cite{macqueen1967some} implemented in scikit-learn~\cite{scikit-learn} on historical incident data provided by the Tennessee Department of Transportation, which consists of $47862$ incidents that occurred from January 2018 to May 2019 in Nashville. We vary the parameter $k$ to divide the total area in consideration into 5, 6, and 7 regions. The cluster centers are initialized uniformly at random from the observed data, and we use the classical expectation-minimization-based iterative approach to compute the final clusters~\cite{scikit-learn}.

\textbf{Surrogate Model:} To learn the surrogate model over waiting times conditional on the number of responders in a region and mean incident arrival rate, we use the random forest regression model~\cite{breiman2001random}. We use the mean squared error (MSE) to measure the quality of a node split, use 150 estimators, and consider $\sqrt{|n|}$ random features for each split, where $n$ is the number of features. The following hyper-parameters were tuned using a grid search: the maximum depth of each tree, the minimum number of observations in a node required to split it, and the minimum number of samples required to be at a leaf node to split its parent.

\definecolor{Gray}{gray}{0.9}
\begin{table}[t]
\caption{Experimental hyper-parameter choices.\label{tab:hyperparams} }
\centering
\begin{tabular}{|l|l|}
\hline
\rowcolor{Gray}
\textbf{Parameter}                                                              & \textbf{Value(s)} \\ \hline
Number of Regions                                                               & \{5, 6, 7\}    \\ \hline
Maximum Time Between Re-Allocations                                             & 60 Minutes     \\ \hline
Incident Service Time                                                           & 20 Minutes     \\ \hline
Responder Speed                                                                 & 30 Mph         \\ \hline
MCTS Iteration Limit                                                            & 1000           \\ \hline
Discount Factor                                                                 & 0.99995        \\ \hline
UCT Tradeoff Parameter $c$                                                          & 1.44           \\ \hline
Number of Generated Incident Samples & 50             \\ \hline
\end{tabular}
\end{table}

\textbf{Stationary incident rates:} We start with a scenario where our forecasting model samples incidents from a Poisson distribution that is stationary (for each cell), meaning that the rate of incident occurrence for each cell is the empirical mean of historical incident occurrence per unit time in the cell. This means that the only utility of the high-level planner in such a case is to divide the overall spatial area into regions and optimize the initial distribution of responders among them. Since the rates are stationary, the initial allocation is maintained throughout the test period under consideration. This scenario lets us test the proposed low-level planning approach in isolation. The experiments were performed on five chains of incidents sampled from the stationary distributions, which have incident counts of \{$939, 937, 974, 1003, 955$\} respectively (for a total of 4808 incidents), and are combined to reduce noise. 


\textbf{Non-stationary incident rates:} We test how our model reacts to changes in incident rates. We identify different types of scenarios that cause the dynamics of spatial temporal incident occurrence and traffic to change in specific areas of Nashville. We look at rush-hour traffic on weekdays (which affects the center of the county), football game days (which affects the area around the football stadium, typically on Saturdays), and Friday evenings (which affects the downtown area). Then, we synthetically simulate spikes in incident rates in the specific areas at times when the areas are expected to see spikes. To further test whether our approach can deal with sudden spikes, we randomly sample the spikes from a Poisson distribution with a rate that varies between two to five times the historical rates of the regions. We create five different trajectories of incidents with varying incident rates, which have incident counts of \{$873, 932, 865, 862, 883$\} respectively (for a total of 4415 incidents). In these experiments we compare using the low-level planner with fixed responder distributions across regions to a full deployment that incorporates the high-level planner to dynamically balance responders across regions. 

\textbf{Responder failures:} An important consideration in emergency response is to quickly account for situations where some responders might be unavailable due to maintenance and breakdowns. We randomly simulate failures of responders lasting 8 hours to understand how our approach deals with such scenarios.

\textbf{Dispatch Policy:} As discussed in section \ref{sec:problem_formulation}, the critical nature of incidents necessitates the use of a \textit{greedy dispatch policy}---if there are any free agents available when an incident is reported, the nearest agent (regardless of region assignment) is dispatched to the incident. If no agents are available, then incidents enter a waiting queue. Once an agent becomes available, if there are any incidents waiting in the queue, it is dispatched to the incident at the front of the queue and that incident is removed from the queue.

\textbf{Baseline Policy:} We compare our approach with a baseline policy that has no responder re-allocation. This baseline emulates current policies in use by cities in which responders are statically assigned to depots and rarely move. The initial responder placement is determined using our proposed high-level policy to ensure all the policies begin with similar responder distributions. The baseline uses the same greedy dispatch policy as our approach.

\section{Results}\label{sec:exp_results}

\begin{figure*}[t]
\centering
\begin{subfigure}{\columnwidth}
  \centering
    \input{figures/pgf/stationary_boxplot}
    \caption{Response time distributions. }
    \label{fig:results_stationary_dist}
\end{subfigure}%
\begin{subfigure}{\columnwidth}
  \centering
  \input{figures/pgf/stationary_average}
  \caption{Average response times.}
  \label{fig:results_stationary_avg}
\end{subfigure}
\caption{Results when applying the baseline and low-level planners to incidents sampled from a stationary rate distribution. Sub-figure (a) presents the full response time distributions; the boxplot represents the data's Inter-Quartile Range (IQR $ = Q_3 - Q_1$), and the whiskers extend to the 9th and 91st percentiles. Sub-figure (b) presents a zoomed in view of the average response times.}
\label{fig:full_results_stationary}
\centering
\begin{subfigure}{.5\textwidth}
  \centering
    \input{figures/pgf/non_stationary_boxplot}
    \caption{Response time distributions. }
    \label{fig:results_non_stationary_dist}
\end{subfigure}%
\begin{subfigure}{.5\textwidth}
  \centering
  \input{figures/pgf/non_stationary_average}
  \caption{Average response times.}
  \label{fig:results_non_stationary_avg}
\end{subfigure}
\caption{Results when applying the baseline, low-level planner (LL Only), and complete hierarchical planner (HL \& LL) when applied to incidents sampled from a non-stationary rate distribution. Sub-figure (a) presents the full response time distributions; the boxplot represents the data's Inter-Quartile Range (IQR $ = Q_3 - Q_1$), and the whiskers extend to the 9th and 91st percentiles. Sub-figure (b) presents a zoomed in view of the average response times.}
\label{fig:full_results_nonstationary}
\end{figure*}

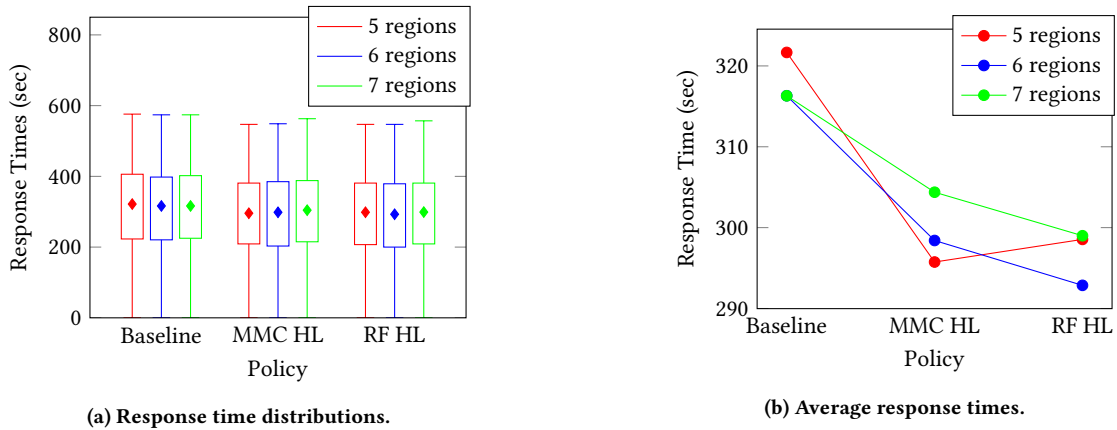
\begin{figure*}[t]
\centering
\begin{subfigure}{.49\textwidth}
  \centering
    \input{figures/pgf/tcps_final_boxplot}
    \caption{Response time distributions. }
    \label{fig:final_dist_full}
\end{subfigure}%
\begin{subfigure}{.49\textwidth}
  \centering
  \input{figures/pgf/tcps_final_average}
  \caption{Average response times.}
  \label{fig:final_avg_full}
\end{subfigure}
\caption{Results when applying the baseline, complete hierarchical planner using the MMC queuing high level planner (MMC HL), and complete planner using the surrogate model high level planner (RF HL) when applied to incidents sampled from a non-stationary rate distribution and using a data-driven travel time router. Sub-figure (a) presents the full response time distributions; the boxplot represents the data's Inter-Quartile Range (IQR $ = Q_3 - Q_1$), and the whiskers extend to the 9\textsuperscript{th} and 91\textsuperscript{st} percentiles. Sub-figure (b) presents a zoomed in view of the average response times.}
\label{fig:full_results_final}
 \end{figure*}

 \begin{figure*}[t]
\centering
\begin{subfigure}{.49\textwidth}
  \centering
    \input{figures/pgf/failure_boxplot}
    \caption{Response time distributions. }
    \label{fig:final_dist_fail}
\end{subfigure}%
\begin{subfigure}{.49\textwidth}
  \centering
  \input{figures/pgf/failure_average}
  \caption{Average response times.}
  \label{fig:final_avg_fail}
\end{subfigure}
\caption{Results when applying the low-level planner only (LL Only), complete hierarchical planner using the MMC queuing high level planner (MMC HL), and complete planner using the surrogate model high level planner (RF HL) when subjected to increasing numbers of simultaneous equipment failures. Sub-figure (a) presents the full response time distributions; the boxplot represents the data's Inter-Quartile Range (IQR $ = Q_3 - Q_1$), and the whiskers extend to the 9th and 91st percentiles. Sub-figure (b) presents a zoomed in view of the average response times.}
\label{fig:full_failure_results}
\end{figure*}
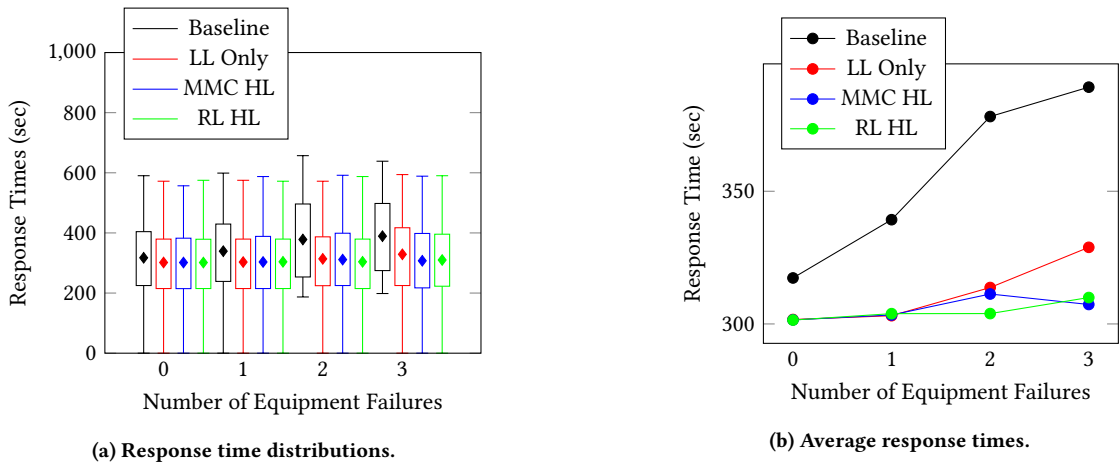

\textbf{Stationary Incident Rates:} We begin by comparing the baseline policy with the proposed low-level planner on incidents sampled from stationary incident rates. Instead of using the data-driven surrogate and travel-time models, we test the low-level planner in isolation, i.e., we use the simpler queue-based model for initial allocations and a travel-time model based on Euclidean distance (we present results with the data-driven models later). The results are shown in figure~\ref{fig:full_results_stationary}. Our first observation is that using the low-level planner reduces response times for all region configurations, improving upon the baseline by \textbf{7.5} seconds on average. This reduction is a significant improvement in the context of emergency response since it is well-known that paramedic response time affects short-term patient survival~\cite{mayer1979emergency}. We also observe a significant shift in the distribution of response times, with the upper quartile of the low-level results being reduced by approximately \textbf{71 seconds} for each region configuration. This reduction in variance indicates that the proposed approach is more consistent than the baseline. As a result, a lesser number of incidents experience large response times. 





\textbf{Non-Stationary Incident Rates:} We now examine the results of experiments using incidents generated from non-stationary incident distributions, which are shown in figure \ref{fig:full_results_nonstationary}. Again, we begin by using the simple queue-based allocation. Our first observation is that response times generally increase relative to the stationary experiments for both the baseline and the proposed approach. This result is expected since the response to incidents sampled from a non-stationary distribution is more challenging to plan for. However, we also observe that our approach can better adapt to the varying rates. The low-level planner in isolation improves upon the baseline's response times by \textbf{18.6 seconds} on average. Introducing the complete hierarchical planner (i.e., both the high-level and low-level planners) improves the result further, reducing response times by \textbf{3 seconds} compared to using only the low-level planner, and \textbf{21.6 seconds} compared to the baseline. We again observe that the region configuration has a small effect on the efficiency of the proposed approach. This result shows that our approach reduces lower response times irrespective of how the original problem is divided into regions. Finally, we also observe that the variance of the response time distributions achieved by the proposed method is not as low as compared to the stationary experiments, which is likely due to the high strain placed on the system from the non-stationary incident rates. 

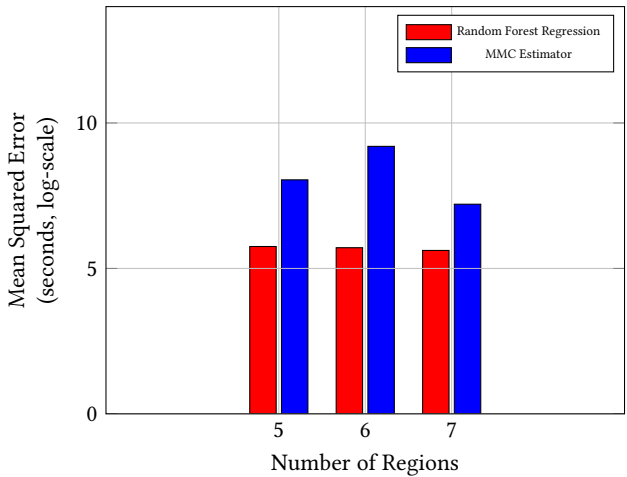
\begin{figure}[t]
    \centering
    \input{figures/pgf/rf_mmc}
    \caption{Mean Squared Error in logarithmic scale for the proposed estimators. The random forest regression model performs significantly better in comparison to the queuing based estimator. However, as the queuing based estimator learns the proportion of wait times among the regions fairly well, it serves as a meaningful heuristic to guide the high-level planner.}
    \label{fig:forecasting_results}
\end{figure}

We now evaluate the surrogate model and its effect on the planner. First, we show how the model performs while forecasting waiting times in unseen test data with respect to the queuing model. We show the results in figure~\ref{fig:forecasting_results}. We observe that the surrogate model based on random forest regression significantly outperforms the queuing-based model; this improvement is expected as the regression model considers travel times and the time taken to drop victims to hospitals through the simulated data. Even though the queuing-based model has large prediction errors, we show that using it as a heuristic to guide the high-level planner outperforms the baseline approach, most likely because such an estimator learns the proportion of waiting times among the regions fairly well.

\begin{figure*}[t]
    \centering
    \includegraphics[width=.8\textwidth]{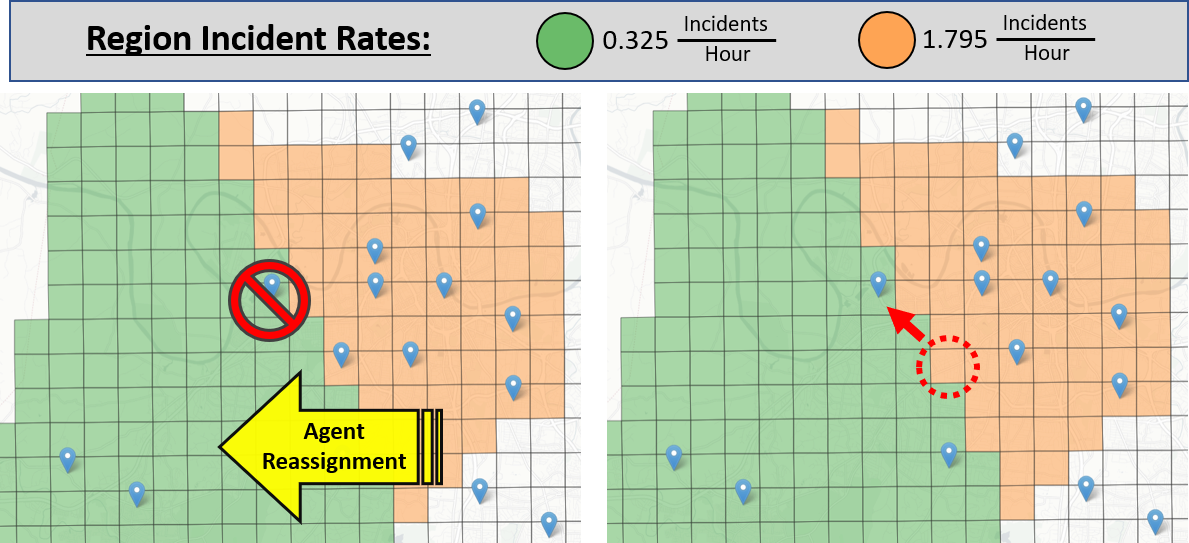}
    \caption{Example of the high-level planner resolving an equipment failure. In sub-figure (left), the agent positioned at the depot marked by the red circle in the green region fails, and the high-level planner determines there is an imbalance across regions. In sub-figure (right), we see the planner move an agent from the depot marked by the red dotted circle to the green region to ensure that the upper left of the region can be serviced. }
    \label{fig:failure}


   
\end{figure*}

Finally, we test the entire hierarchical planning pipeline (with both the surrogate model and the queuing-based model for initial allocation) and compare it with the baseline approach. We also use the data-driven travel time router to replicate realistic travel times. We present the results in figure \ref{fig:full_results_final}. We see that the hierarchical planner with the data-driven surrogate model usually outperforms the other approaches. On average, it improves response times by about \textbf{23 seconds} with respect to the baseline model and by about \textbf{6 seconds} with respect to a hierarchical planner that uses a queuing model for the high-level planner when using each model's best region segmentation. 
We note that the high-level planner using the queuing model outperforms the surrogate model in one case (5 regions). A potential cause for this might be shifts in the underlying environmental distributions, which have been shown to cause learning-based approaches to perform poorly in other domains~\cite{park2020calibrated, laine2020eyes, ramakrishna2021efficient}. Future analyses are needed to examine the impact of such shifts on the surrogate model and identify other potential causes for the queuing model to perform better in some situations.

\textbf{Responder Failures: } Results on the non-stationary incident distribution demonstrate the effectiveness of the hierarchical planner when there are shifts in the spatial distribution of incidents. We now examine its response to equipment failures within the ERM system. Figure \ref{fig:failure} illustrates an example (from our experiments) of how the planner can adapt to equipment failures. When a responder in the green region fails, the high-level planner determines that imbalance in the spatial distribution of the responders. Intuitively, due to the failure incidents occurring in the upper left cells of the green region could face long response times. Therefore, the planner reallocates a responder from the orange region to the green region. To examine how equipment failures impact the proposed approach, we simulated several responder failures and compared system performance using the baseline policy, the low-level planner in isolation, the full hierarchical approach using the MMC queue high-level planner, and the full approach using the surrogate model. We show the results in figure \ref{fig:full_failure_results}. Naturally, as the number of failures increases, response times increase with fewer responders. However, we observe that the proposed hierarchical approach intelligently allocates the remaining responders to outperform the baseline and low-level planner in isolation. Indeed, when there are three simultaneous failures, using the MMC queuing-based hierarchical planner improves response times by about \textbf{82 seconds} compared to the baseline policy and about \textbf{22 seconds} compared to using only the low-level planner.

\textbf{Allocation Computation Times: } 
Decisions using the proposed approach take  \textbf{180.29} seconds on average. Note that this is the time that our system takes to optimize the allocation of responders. Dispatch decisions are greedy and occur instantaneously. Hence, our system can easily be used by first responders on the field without hampering existing operational speed.


%% file: figures/pgf/stationary_boxplot.tex
\def\boxwidth{0.75}%

\begin{tikzpicture}
\begin{axis}[
    scale only axis,
    axis on top,
    xlabel={Policy},
    ymin=0, ymax=500,
    ylabel={Response Times (sec)},
    boxplot/draw direction=y,
    xtick={2, 6},
    xticklabels={Baseline, LL Only},
    tick align=inside,
    xtick style={draw=none},
    cycle list={{red},{blue},{green}},
    height=4cm,
    width=5cm,
    /pgfplots/line legend with two nodes/.style 2 args={
    \draw[##1, draw=no markers] plot coordinates {
        (0.6cm,0cm)
        (0.6cm,0cm)
        (0.6cm,0cm)
      }
      node[pos=0,#1]{}
      node[#2]{};
    }, 
    legend style={at={(0.58,.87)},anchor=west}
]
\addlegendimage{no markers,red}
\addlegendentry{5 regions};
\addlegendimage{no markers,blue}
\addlegendentry{6 regions};
\addlegendimage{no markers,green}
\addlegendentry{7 regions};
\addplot+[
    boxplot prepared={draw position=1,
        lower whisker=0,
        lower quartile=120,
        average=162.3744895,
        upper quartile=240,
        upper whisker=268.3281572,
        box extend=\boxwidth,
    },
]
coordinates{};

\addplot+[
    boxplot prepared={draw position=2,
        lower whisker=0,
        lower quartile=120,
        average=165.4151597,
        upper quartile=240,
        upper whisker=339.4112549,
        box extend=\boxwidth,
    },
]
coordinates{};

\addplot+[
    boxplot prepared={draw position=3,
        lower whisker=0,
        lower quartile=120,
        average=164.2713039,
        upper quartile=240,
        upper whisker=339.4112549,
        box extend=\boxwidth,
    },
]
coordinates{};

\addplot+[
    boxplot prepared={draw position=5,
        lower whisker=0,
        lower quartile=120,
        average=156.6882741,
        upper quartile=169.7056274,
        upper whisker=268.3281572,
        box extend=\boxwidth,
    },
]
coordinates{};

\addplot+[
    boxplot prepared={draw position=6,
        lower whisker=0,
        lower quartile=120,
        average=155.604194,
    upper quartile=169.7056274,
        upper whisker=268.3281572,
        box extend=\boxwidth,
    },
]
coordinates{};

\addplot+[
    boxplot prepared={draw position=7,
        lower whisker=0,
        lower quartile=120,
        average=157.5305,
        upper quartile=169.7056274,
        upper whisker=268.3281572,
        box extend=\boxwidth,
    },
]
coordinates{};
\end{axis}
\end{tikzpicture}

%% file: figures/pgf/stationary_average.tex
\begin{tikzpicture}
\begin{axis}[
	xlabel=Policy,
	ymin=154, ymax=167,
	ylabel=Response Time (sec),
	width=6.3cm,height=5.3cm,
	xmin=0, xmax=7,
	xtick={2, 5},
    xticklabels={Baseline, LL Only},
    tick align=inside,
    xtick style={draw=none},
    legend style={at={(0.55,.87)},anchor=west}
    ]
\addplot[color=red,mark=*] coordinates {
	(2, 162.3744895)
	(5, 156.6882741)
};

\addplot[color=blue,mark=*] coordinates {
	(2, 165.415159666845)
	(5, 155.604194245378)
};

\addplot[color=green,mark=*] coordinates {
	(2, 164.2713039)
	(5, 157.530529)
};
\legend{5 regions, 6 regions, 7 regions}
\end{axis}
\end{tikzpicture}


%% file: figures/pgf/non_stationary_boxplot.tex
\def\boxwidth{0.75}%

\begin{tikzpicture}
\begin{axis}[
    scale only axis,
    axis on top,
    ymin=0, ymax=550,
    ylabel={Response Times (sec)},
    boxplot/draw direction=y,
    xtick={2, 6, 10},
    xticklabels={Baseline, LL Only, HL \& LL},
    xlabel={Policy},
    tick align=inside,
    xtick style={draw=none},
    cycle list={{red},{blue},{green}},
    height=4cm,
    width=5cm,
    /pgfplots/line legend with two nodes/.style 2 args={
    \draw[##1, draw=no markers] plot coordinates {
        (0.6cm,0cm)
        (0.6cm,0cm)
        (0.6cm,0cm)
      }
      node[pos=0,#1]{}
      node[#2]{};
    },
    legend style={at={(0.58,.87)},anchor=west}
]
\addlegendimage{no markers,red}
\addlegendentry{5 regions};
\addlegendimage{no markers,blue}
\addlegendentry{6 regions};
\addlegendimage{no markers,green}
\addlegendentry{7 regions};
\addplot+[
    boxplot prepared={draw position=1,
        lower whisker=0,
        lower quartile=120,
        average=185.8822214,
        upper quartile=240,
        upper whisker=379.473319,
        box extend=\boxwidth,
    },
]
coordinates{};

\addplot+[
    boxplot prepared={draw position=2,
        lower whisker=0,
        lower quartile=120,
        average=191.9680626,
        upper quartile=268.3281572,
        upper whisker=379.4733193,
        box extend=\boxwidth,
    },
]
coordinates{};

\addplot+[
    boxplot prepared={draw position=3,
        lower whisker=0,
        lower quartile=120,
        average=191.6035288,
        upper quartile=268.3281572,
        upper whisker=379.4733193,
        box extend=\boxwidth,
    },
]
coordinates{};

\addplot+[
    boxplot prepared={draw position=5,
        lower whisker=0,
        lower quartile=120,
        average=169.174178,
        upper quartile=240,
        upper whisker=339.4112549,
        box extend=\boxwidth,
    },
]
coordinates{};

\addplot+[
    boxplot prepared={draw position=6,
        lower whisker=0,
        lower quartile=120,
        average=170.9151886,
    upper quartile=240,
        upper whisker=360,
        box extend=\boxwidth,
    },
]
coordinates{};

\addplot+[
    boxplot prepared={draw position=7,
        lower whisker=0,
        lower quartile=120,
        average=173.5392086,
        upper quartile=240,
        upper whisker=360,
        box extend=\boxwidth,
    },
]
coordinates{};

\addplot+[
    boxplot prepared={draw position=9,
        lower whisker=0,
        lower quartile=120,
        average=166.4264318,
        upper quartile=240,
        upper whisker=339.4112549,
        box extend=\boxwidth,
    },
]
coordinates{};

\addplot+[
    boxplot prepared={draw position=10,
        lower whisker=0,
        lower quartile=120,
        average=168.1263735,
    upper quartile=240,
        upper whisker=354.6469263,
        box extend=\boxwidth,
    },
]
coordinates{};

\addplot+[
    boxplot prepared={draw position=11,
        lower whisker=0,
        lower quartile=120,
        average=170.322567,
        upper quartile=240,
        upper whisker=360,
        box extend=\boxwidth,
    },
]
coordinates{};
\end{axis}
\end{tikzpicture}

%% file: figures/pgf/non_stationary_average.tex
\begin{tikzpicture}
\begin{axis}[
	xlabel=Policy,
	ylabel=Response Time (sec),
	width=6.3cm,height=5.3cm,
	xtick={2, 5, 8},
    xticklabels={Baseline, LL Only, HL \& LL},
    tick align=inside,
    xtick style={draw=none},
    legend style={at={(0.55,.87)},anchor=west}
]
\addplot[color=red,mark=*] coordinates {
	(2, 185.8822214)
	(5, 169.174178)
	(8, 166.4264318)
};

\addplot[color=blue,mark=*] coordinates {
	(2, 191.9680626)
	(5, 170.9151886)
	(8, 168.1263735)
};

\addplot[color=green,mark=*] coordinates {
	(2, 191.6035288)
	(5, 173.5392086)
	(8, 170.322567)
};
\legend{5 regions, 6 regions, 7 regions}
\end{axis}
\end{tikzpicture}


%% file: figures/pgf/tcps_final_boxplot.tex
\def\boxwidth{0.75}%

\begin{tikzpicture}
\begin{axis}[
    scale only axis,
    axis on top,
    ymin=0, ymax=850,
    ylabel={Response Times (sec)},
    boxplot/draw direction=y,
    xtick={2, 6, 10},
    xticklabels={Baseline, MMC HL, RF HL},
    xlabel={Policy},
    tick align=inside,
    xtick style={draw=none},
    cycle list={{red},{blue},{green}},
    height=4cm,
    width=5cm,
    /pgfplots/line legend with two nodes/.style 2 args={
    \draw[##1, draw=no markers] plot coordinates {
        (0.6cm,0cm)
        (0.6cm,0cm)
        (0.6cm,0cm)
      }
      node[pos=0,#1]{}
      node[#2]{};
    },
    legend style={at={(0.58,.87)},anchor=west}
]
\addlegendimage{no markers,red}
\addlegendentry{5 regions};
\addlegendimage{no markers,blue}
\addlegendentry{6 regions};
\addlegendimage{no markers,green}
\addlegendentry{7 regions};
\addplot+[
    boxplot prepared={draw position=1,
        lower whisker=0,
        lower quartile=223,
        average=321.66704416761,
        upper quartile=406,
        upper whisker=576,
        box extend=\boxwidth,
    },
]
coordinates{};

\addplot+[
    boxplot prepared={draw position=2,
        lower whisker=0,
        lower quartile=220.5,
        average=316.31053227633,
        upper quartile=398,
        upper whisker=574,
        box extend=\boxwidth,
    },
]
coordinates{};

\addplot+[
    boxplot prepared={draw position=3,
        lower whisker=0,
        lower quartile=225,
        average=316.303057757644,
        upper quartile=402,
        upper whisker=574,
        box extend=\boxwidth,
    },
]
coordinates{};

\addplot+[
    boxplot prepared={draw position=5,
        lower whisker=0,
        lower quartile=209,
        average=295.7449603624,
        upper quartile=381,
        upper whisker=547,
        box extend=\boxwidth,
    },
]
coordinates{};

\addplot+[
    boxplot prepared={draw position=6,
        lower whisker=0,
        lower quartile=203,
        average=298.408154020385,
    upper quartile=385,
        upper whisker=548.74,
        box extend=\boxwidth,
    },
]
coordinates{};

\addplot+[
    boxplot prepared={draw position=7,
        lower whisker=0,
        lower quartile=215,
        average=304.379614949037,
        upper quartile=388,
        upper whisker=563,
        box extend=\boxwidth,
    },
]
coordinates{};

\addplot+[
    boxplot prepared={draw position=9,
        lower whisker=0,
        lower quartile=207,
        average=298.543884892086,
        upper quartile=381,
        upper whisker=547,
        box extend=\boxwidth,
    },
]
coordinates{};

\addplot+[
    boxplot prepared={draw position=10,
        lower whisker=0,
        lower quartile=200,
        average=292.861151079136,
    upper quartile=379,
        upper whisker=547,
        box extend=\boxwidth,
    },
]
coordinates{};

\addplot+[
    boxplot prepared={draw position=11,
        lower whisker=0,
        lower quartile=209,
        average=298.997281993205,
        upper quartile=381,
        upper whisker=557,
        box extend=\boxwidth,
    },
]
coordinates{};
\end{axis}
\end{tikzpicture}

%% file: figures/pgf/tcps_final_average.tex
\begin{tikzpicture}
\begin{axis}[
	xlabel=Policy,
	ylabel=Response Time (sec),
	width=6.3cm,height=5.3cm,
	xtick={2, 5, 8},
    xticklabels={Baseline, MMC HL, RF HL},
    tick align=inside,
    xtick style={draw=none},
    legend style={at={(0.55,.87)},anchor=west}
    ]
\addplot[color=red,mark=*] coordinates {
	(2, 321.6670442)
	(5, 295.7449604)
	(8, 298.5438849)
};

\addplot[color=blue,mark=*] coordinates {
	(2, 316.3105323)
	(5, 298.408154)
	(8, 292.8611511)
};

\addplot[color=green,mark=*] coordinates {
	(2, 316.30305775764)
	(5, 304.379614949037)
	(8, 298.997282)
};
\legend{5 regions, 6 regions, 7 regions}
\end{axis}
\end{tikzpicture}


%% file: figures/pgf/failure_boxplot.tex
\def\boxwidth{0.75}%

\begin{tikzpicture}
\begin{axis}[
    scale only axis,
    axis on top,
    ymin=0, ymax=1000,
    ylabel={Response Times (sec)},
    boxplot/draw direction=y,
    xtick={2, 6, 10, 14},
    xticklabels={0, 1, 2, 3},
    xlabel={Number of Equipment Failures},
    tick align=inside,
    xtick style={draw=none},
    cycle list={{black},{red},{blue},{green}},
    height=4cm,
    width=5cm,
    /pgfplots/line legend with two nodes/.style 2 args={
    \draw[##1, draw=no markers] plot coordinates {
        (0.6cm,0cm)
        (0.6cm,0cm)
        (0.6cm,0cm)
        (0.6cm,0cm)
      }
      node[pos=0,#1]{}
      node[#2]{};
    },
    legend style={at={(0.05,.93)},anchor=west}
]
\addlegendimage{no markers,black}
\addlegendentry{Baseline};
\addlegendimage{no markers,red}
\addlegendentry{LL Only};
\addlegendimage{no markers,blue}
\addlegendentry{MMC HL};
\addlegendimage{no markers,green}
\addlegendentry{RL HL};
\addplot+[
    boxplot prepared={draw position=1,
        lower whisker=0,
        lower quartile=225,
        average=317.322580645161,
        upper quartile=404.25,
        upper whisker=590.31,
        box extend=\boxwidth,
    },
]
coordinates{};
\addplot+[
    boxplot prepared={draw position=2,
        lower whisker=0,
        lower quartile=215,
        average=301.673387096774,
        upper quartile=379.5,
        upper whisker=572,
        box extend=\boxwidth,
    },
]
coordinates{};

\addplot+[
    boxplot prepared={draw position=3,
        lower whisker=0,
        lower quartile=214.75,
        average=301.524193548387,
        upper quartile=382.5,
        upper whisker=556.77,
        box extend=\boxwidth,
    },
]
coordinates{};

\addplot+[
    boxplot prepared={draw position=4,
        lower whisker=0,
        lower quartile=215,
        average=301.463709677419,
        upper quartile=379,
        upper whisker=575.08,
        box extend=\boxwidth,
    },
]
coordinates{};

\addplot+[
    boxplot prepared={draw position=5,
        lower whisker=0,
        lower quartile=238.75,
        average=339.282258064516,
        upper quartile=429.5,
        upper whisker=599,
        box extend=\boxwidth,
    },
]
coordinates{};
\addplot+[
    boxplot prepared={draw position=6,
        lower whisker=0,
        lower quartile=215,
        average=303.104838709677,
        upper quartile=379.5,
        upper whisker=575.08,
        box extend=\boxwidth,
    },
]
coordinates{};

\addplot+[
    boxplot prepared={draw position=7,
        lower whisker=0,
        lower quartile=215,
        average=303.375,
    upper quartile=388.5,
        upper whisker=587.55,
        box extend=\boxwidth,
    },
]
coordinates{};

\addplot+[
    boxplot prepared={draw position=8,
        lower whisker=0,
        lower quartile=215,
        average=303.879032258064,
        upper quartile=379.5,
        upper whisker=572,
        box extend=\boxwidth,
    },
]
coordinates{};

\addplot+[
    boxplot prepared={draw position=9,
        lower whisker=187,
        lower quartile=253.5,
        average=378.145161290322,
        upper quartile=496.25,
        upper whisker=657.01,
        box extend=\boxwidth,
    },
]
coordinates{};
\addplot+[
    boxplot prepared={draw position=10,
        lower whisker=0,
        lower quartile=224.5,
        average=313.721774193548,
        upper quartile=387,
        upper whisker=572,
        box extend=\boxwidth,
    },
]
coordinates{};

\addplot+[
    boxplot prepared={draw position=11,
        lower whisker=0,
        lower quartile=225,
        average=311.278225806451,
        upper quartile=399,
        upper whisker=591.77,
        box extend=\boxwidth,
    },
]
coordinates{};

\addplot+[
    boxplot prepared={draw position=12,
        lower whisker=0,
        lower quartile=215,
        average=303.903225806451,
        upper quartile=379.5,
        upper whisker=587.55,
        box extend=\boxwidth,
    },
]
coordinates{};

\addplot+[
    boxplot prepared={draw position=13,
        lower whisker=198.46,
        lower quartile=274.75,
        average=389.221774193548,
        upper quartile=498,
        upper whisker=638.54,
        box extend=\boxwidth,
    },
]
coordinates{};
\addplot+[
    boxplot prepared={draw position=14,
        lower whisker=0,
        lower quartile=225,
        average=328.870967741935,
        upper quartile=417.25,
        upper whisker=594.08,
        box extend=\boxwidth,
    },
]
coordinates{};

\addplot+[
    boxplot prepared={draw position=15,
        lower whisker=0,
        lower quartile=217.25,
        average=307.334677419354,
        upper quartile=398.25,
        upper whisker=588.77,
        box extend=\boxwidth,
    },
]
coordinates{};

\addplot+[
    boxplot prepared={draw position=16,
        lower whisker=0,
        lower quartile=222.75,
        average=309.951612903225,
        upper quartile=395.75,
        upper whisker=590.31,
        box extend=\boxwidth,
    },
]
coordinates{};
\end{axis}
\end{tikzpicture}

%% file: figures/pgf/failure_average.tex
\begin{tikzpicture}
\begin{axis}[
	xlabel=Number of Equipment Failures,
	ylabel=Response Time (sec),
	width=6.3cm,height=5.3cm,
	xtick={2, 5, 8, 11},
    xticklabels={0,1,2,3},
    tick align=inside,
    xtick style={draw=none},
    legend style={at={(0.05,.93)},anchor=west}
    ]
\addplot[color=black,mark=*] coordinates {
	(2, 317.322580645161)
	(5, 339.282258064516)
	(8, 378.145161290322)
	(11, 389.221774193548)
};    

\addplot[color=red,mark=*] coordinates {
	(2, 301.673387096774)
	(5, 303.104838709677)
	(8, 313.721774193548)
	(11, 328.870967741935)
};

\addplot[color=blue,mark=*] coordinates {
	(2, 301.524193548387)
	(5, 303.375)
	(8, 311.278225806451)
	(11, 307.334677419354)
};

\addplot[color=green,mark=*] coordinates {
	(2, 301.463709677419)
	(5, 303.879032258064)
	(8, 303.903225806451)
	(11, 309.951612903225)
};
\legend{Baseline, LL Only, MMC HL, RL HL}
\end{axis}
\end{tikzpicture}

%% file: figures/pgf/rf_mmc.tex
\begin{tikzpicture}
\begin{axis}[
    ybar,
    enlarge x limits=1,
    axis on top,
    ymin=0, ymax=14, area legend,
    legend style={font=\tiny},
     grid=both,
    grid style={line width=.1pt, draw=gray!10},
    major grid style={line width=.2pt,draw=gray!50},
    ylabel style={align=center},
    ylabel={Mean Squared Error\\(seconds, log-scale)},
    symbolic x coords={5,6,7},
    xtick = data,
    xlabel={Number of Regions},
    tick align=inside,
    x tick label style={ align=center},
    xtick style={draw=none},
    height=7cm,
    width=\columnwidth,
    ]
\addplot[fill=red] coordinates {(5,5.753583908) (6,5.711951437) (7,5.618112231)};
\addplot[fill=blue] coordinates {(5,8.044824468) (6,9.196409914) (7,7.207722728)};
\addlegendimage{fill,red}
\addlegendentry{Random Forest Regression};
\addlegendimage{fill,blue}
\addlegendentry{MMC Estimator};
\end{axis}
\end{tikzpicture}

%% file: related_work.tex
\section{Related work}\label{sec:related}

Markov decision processes can be directly solved using dynamic programming when the transition dynamics of the system are known ~\cite{kochenderfer2015decision}. The transition dynamics are typically unknown for resource allocation problems in complex environments like urban areas,~\cite{mukhopadhyay2020review}. The Simulate-and-Transform (\textit{SimTrans}) algorithm \cite{mukhopadhyayAAMAS18} can be used to address this issue; it performs canonical policy iteration with an added computation. In order to estimate values (utilities) of states, the algorithm simulates the entire system of incident occurrence and responder dispatch and keeps track of all states, transitions, and actions, and gradually builds statistically confident estimates of the transition probabilities.

While \textit{SimTrans} finds a close approximation of the optimal policy (assuming that the estimates of the transition probabilities are close to the true probabilities), the process is extremely slow and unsuited to dynamic environments. As an example, even if a single agent (ambulance in this case) breaks down, the entire process of estimating transition probabilities and learning a policy must be repeated. To better react to dynamic environmental conditions, decentralized and online approaches have been explored~\cite{claes2017decentralised,MukhopadhyayICCPS}. For example, ~\citeauthor{claes2017decentralised}~\cite{claes2017decentralised} entrust each agent to build its own decision tree and show how computationally cheap models can be used by agents to estimate the actions of other agents as the trees are built.

An orthogonal approach to solve large-scale MDPs is using hierarchical planning~\cite{hauskrecht2013hierarchical}. Such an approach focuses on learning local policies, known as \textit{macros}, over subsets of the state space. The concept of macro-actions was introduced separately from hierarchical planning as means to reuse a learned mapping from states to actions to solve multiple MDPs when objectives change~\cite{sutton1995td, precup1998multi}. Later, the macro-policies were used in hierarchical models to address the issue of large state and action spaces~\cite{forestier1978multilayer,hauskrecht2013hierarchical}. 

We also describe how allocation and dispatch are handled in emergency response. First, note that the distinction between allocation and response problems can be hazy since any solution to the allocation problem implicitly creates a policy for response (greedy response based on the allocation)~\cite{mukhopadhyay2020review}. We use a similar approach in this paper because greedy response satisfies the constraints under which first responders operate. A common metric for optimizing resource allocation is coverage~\cite{toregas_location_1971,church1974maximal, gendreau_solving_1997}. Waiting time constraints are often used as constraints in approaches that maximize coverage~\cite{silva2008locating,mukhopadhyayAAMAS17}. Decision-theoretic models have also been widely used to design ERM systems. For example, \citeauthor{keneally2016markov}~\cite{keneally2016markov} model the resource allocation and dispatch problem in ERM as a continuous-time MDP, while we have previously used a semi-Markovian process~\cite{mukhopadhyayAAMAS18}. Allocation in ERM can also be addressed by optimizing the distance between facilities and demand locations~\cite{dzator_effective_2013, MukhopadhyayICCPS}, and explicitly optimizing for patient survival~\cite{erkut_ambulance_2008, knight_ambulance_2012}.

%% file: Conclusion.tex
\section{Conclusion} \label{sec:conclusion}

We present a hierarchical planning approach for dynamic resource allocation in city scale cyber-physical system (CPS). We formulate a general decision-theoretic problem for that can be used in a variety of resource allocation settings. We model the overall problem as a Multi-Agent Semi-Markov Decision Process (MSMDP), and show how to leverage the problem's spatial structure to decompose the MSMDP into smaller and tractable sub-problems. We then detail how a hierarchical planner can employ a low-level planner to solve these sub-problems, while a high-level planner identifies situations in which resources must be moved across region lines. We use emergency response as a case-study and validate the proposed approach with data from a major metropolitan area in the USA. Our experiments show that our proposed hierarchical approach offers significant improvements when compared to the state-of-the-art in emergency response planning, as it maintains system fairness while significantly decreasing average incident response times. We also find that it is robust to equipment failure and is computationally efficient enough to be deployed in the field without hampering existing operational speed. While this work demonstrates the potential of hierarchical decision making, several non-trivial technical challenges remain, including how to optimally divide a spatial area such that the solutions of the sub-problems maximize the overall utility of the original problem. We will explore these challenges in future work.